\documentclass[
  aps,
  prx,
  twocolumn,
  superscriptaddress,
  amsmath,amssymb,
  longbibliography
]{revtex4-2}

\usepackage[utf8]{inputenc}
\usepackage{amsthm,bm}
\usepackage{mathtools,mathdots}
\usepackage{pifont}
\usepackage[dvipdfmx]{graphicx}
\usepackage{xurl}
\usepackage{comment}
\usepackage{placeins}
\usepackage{xcolor}

\begin{document}

\title{Stochastic Thermodynamics for Autoregressive Generative Models: A Non-Markovian Perspective}

\author{Takahiro Sagawa}
\affiliation{Department of Applied Physics, The University of Tokyo, Bunkyo-ku, Tokyo 113-8656, Japan}
\affiliation{Quantum-Phase Electronics Center (QPEC), The University of Tokyo, Bunkyo-ku, Tokyo 113-8656, Japan}
\affiliation{Inamori Research Institute for Science (InaRIS), Kyoto-shi, Kyoto 600-8411, Japan}

\date{\today}

\begin{abstract}
Autoregressive generative models --- including Transformers, recurrent neural networks, classical Kalman filters, state space models, and Mamba  --- all generate sequences by sampling each output from a deterministic summary of the past, producing genuinely non-Markovian observed processes.  We develop a general theoretical framework based on stochastic thermodynamics for this class of architectures and introduce the entropy production, which can be efficiently estimated from sampled trajectories without exponential overhead, despite the non-Markovian nature of the observed dynamics.  
As a proof-of-concept experiment with a large language model (LLM), we evaluate the entropy production for  a pre-trained Transformer-based model, GPT-2.
We find that the token-level entropy production is dominated by a syntactic artifact, while the sentence-level entropy production tends to be larger for causally ordered than for non-causal text sets. This observation is supported by a re-evaluation with a substantially larger model, Qwen3-4B-Base.
We also demonstrate the framework in the linear Gaussian case, where the model reduces to the Kalman innovation representation and the entropy production admits an analytical expression.  
We also show that the entropy production decomposes exactly into non-negative per-step contributions in terms of retrospective inference, and each of those terms further splits into information-theoretically meaningful terms:  a compression loss and a model mismatch.  Our results establish a  bridge between stochastic thermodynamics and modern generative models, and provide a starting point
for using irreversibility as a quantitative probe of the highly non-Markovian processes generated by models such as LLMs.
\end{abstract}

\maketitle

%======================================================================
\section{Introduction}
%======================================================================

Stochastic thermodynamics provides a general framework for quantifying
irreversibility in stochastic processes, with entropy production
playing the role of a central diagnostic
\cite{Seifert2012,Seifert2005,Jarzynski1997,Crooks1999,PelitiPigolotti2021}.
Although the theory is most fully developed for Markovian processes,
substantial progress has extended it  in several
complementary directions.
One line of work characterizes irreversibility directly from the forward and backward path measures, including for non-Markovian time series~\cite{RoldanParrondo2010,RoldanParrondo2012,SearaMachtaMurrell2021,GomezMarinParrondoVandenBroeck2008,KawaguchiNakayama2013,KahlenEhrich2018,CrooksStill2019,SpeckSeifert2007,OhkumaOhta2007}. A related line of work studies non-Markovian or partially observed entropy production through various constructions~\cite{Esposito2012,RahavJarzynski2007,PuglisiPigolottiRondoniVulpiani2010,PolettiniEsposito2017,KanazawaDechant2025,VanDerMeerDeguntherSeifert2023,DeguntherVanDerMeerSeifert2024,MunakataRosinberg2014,RosinbergMunakataTarjus2015}.
In parallel, the interplay among thermodynamics, information, and computation has been
extensively studied~\cite{SagawaUeda2010,SagawaUeda2012,SagawaUeda2012b,StillSivakBellCrooks2012,HorowitzEsposito2014,ItoSagawa2013,ParrondoHorowitzSagawa2015,Ito2016BTE,WolpertKolchinsky2020,Manzano2021,ManzanoKardesRoldanWolpert2024,WolpertKorbelLynn2024}.
Stochastic thermodynamics of neural networks has also been explored, including in the context of the learning process~\cite{GoldtSeifert2017}, and more recently for dense associative memory (modern Hopfield) networks~\cite{RookeKrotovBalasubramanianWolpert2026}.

Autoregressive generative models have become the
central architecture of modern generative AI.
These models generate a sequence of elements,
with each new element drawn from a conditional distribution that depends on a
deterministic summary of the preceding observations.
For a language model, each such element is a token: a discrete unit such as a word, a subword, or a punctuation mark.
The Transformer architecture \cite{Vaswani2017,Radford2019,Brown2020,Ramsauer2021},
which underlies most contemporary large language models (LLMs),
uses causal self-attention to construct a context vector from the full
past sequence; the deterministic summary cannot in general be reduced to a
fixed-order recursive update,
so the resulting observed token sequence is genuinely non-Markovian.
Recurrent neural networks (RNNs)
\cite{Elman1990,HochreiterSchmidhuber1997} are also regarded as autoregressive generative models, but have a simpler structure with recursive updates of latent states.
The Kalman filter \cite{Kalman1960,AndersonMoore1979,LindquistPicci2015}, long studied in
control theory and signal processing, can be viewed as a generative model of this recursive class.
Moreover, structured state space models (SSMs) and Mamba (an alternative architecture for LLMs)
\cite{GuGoel2022,GuDao2023,SmithWarringtonLinderman2023} also fall into this recursive class.
Despite their diverse origins, all of these architectures share the same
abstract structure: a stochastic emission from a deterministic latent state
that encodes past observations.

In this paper, we develop a stochastic thermodynamic framework for
the class of non-Markovian processes generated by autoregressive
models with deterministic internal memory.
This class encompasses diverse architectures as described above;
we bring them under a single
 framework (Table~\ref{tab:examples}) and 
develop the stochastic thermodynamics of the resulting
non-Markovian observed process in a unified manner,
by constructing a backward process with the same
architectural components applied in reversed temporal order.

Conceptually, our definition of the entropy production builds on a line of research in which the KL divergence between forward and backward path measures serves as an operational characterization of irreversibility for the observed process~\cite{RoldanParrondo2010,RoldanParrondo2012,KahlenEhrich2018,SearaMachtaMurrell2021,GomezMarinParrondoVandenBroeck2008,KawaguchiNakayama2013,CrooksStill2019,SpeckSeifert2007,OhkumaOhta2007}.
The central observation about the autoregressive class studied here is that every factor in the forward/backward path-probability ratio is supplied explicitly by the emission kernel evaluated at a deterministic latent state, making the stochastic entropy production computable from a single sampled trajectory
despite genuine non-Markovianity.

As a proof-of-concept experiment with a pre-trained LLM, we evaluate the stochastic entropy production for GPT-2. We show that the token-level entropy production is dominated by a syntactic artifact
of token-order reversal, whereas the block-level coarse-grained entropy production,
which reverses the order of sentences rather than individual tokens, tends to be larger for causally ordered than for non-causal texts.
Additionally, a re-evaluation is performed with a substantially larger model, Qwen3-4B-Base, which reproduces the qualitative pattern with improved statistical resolution.

LLMs are often argued to acquire internal representations of the real world, including spatial and temporal structure --- a capacity sometimes discussed in terms of a learned \textit{world model}~\cite{LiHopkinsBauViegasPfisterWattenberg2023,GurneeTegmark2024}. From this perspective, the entropy production of an LLM ultimately aims to probe the irreversibility of the real-world process described by the token sequence --- ideally, the temporal or causal ordering of the events narrated in the text.
Our GPT-2 and Qwen3-4B-Base experiments should be viewed as a first step toward assessing this possibility.

Meanwhile, as an analytically tractable demonstration, we examine the linear
Gaussian case, where the autoregressive model coincides with the
innovation representation of the Kalman filter.
We derive an analytical expression for the entropy production, which is numerically verified by applying the
Monte Carlo sampling procedure.

Furthermore, we show that the entropy production $\mathcal{S}_y$
admits an exact \emph{retrospective} decomposition into a sum of non-negative per-step
contributions $\mathcal{D}_t \geq 0$, each measuring how well the
backward model retrodicts the present observation $y_t$ from the
future.
Each $\mathcal{D}_t$ further splits into a \emph{compression loss}
$\mathcal{L}_t$, arising when the backward latent state is a lossy
summary of the future, and a \emph{model mismatch}
$\mathcal{M}_t$, arising because the emission kernel designed for
forward prediction is reused in the backward direction.
The decomposition is formally reminiscent of the evidence lower bound
(ELBO) decompositions used in variational inference
\cite{KingmaWelling2014,HoffmanJohnson2016,AlemiFischerDillonMurphy2018},
but arises from a fundamentally different starting point (namely, time
reversal and entropy production).
This connection suggests that the stochastic-thermodynamic and machine-learning
perspectives on generative models may benefit from further mutual exchange.

This paper is organized as follows.
In Section~\ref{sec:setup}, we formulate the general autoregressive framework with deterministic latent memory.
In Section~\ref{sec:backward_ep}, we construct the backward process by reusing the architectural components in reversed temporal order and define the entropy production as the KL divergence between the forward and backward path measures.
In Section~\ref{sec:sampling}, we analyze the computational cost of estimating the entropy production, show that the autoregressive structure renders it tractable without exponential sampling overhead, and introduce a temporal coarse-graining.
In Section~\ref{sec:gpt2-demo}, we present a proof-of-concept experiment with GPT-2,   evaluating both the token-level and block-level entropy production.
In Section~\ref{sec:linear-gaussian-example}, we illustrate the framework in the linear Gaussian case, where the model reduces to the Kalman innovation representation and the entropy production admits an analytical expression.
In Section~\ref{sec:retro}, we derive an exact decomposition of the entropy production into non-negative per-step contributions, each further splitting into a compression loss and a model mismatch, and discuss the connection to the thermodynamics of information.
We conclude with a summary and perspectives in Section~\ref{sec:summary}.

%======================================================================
\section{Setup}
\label{sec:setup}
%======================================================================
In this section, we introduce a general framework for autoregressive generative models with deterministic latent memory, and show that Transformers, RNNs, Kalman filters, SSMs, and Mamba all fall into this class  (Table~\ref{tab:examples}).
While each of these concrete architectures is well known, establishing such a unified framework itself constitutes one of the contributions of this paper.

\subsection{State variables and forward process}
\label{subsec:setup}

Consider the following stochastic process.
The variables are $y_t$ ($t=1,2,\dots,T$) and $h_t$ ($t=0,1,\dots,T$), where $h_0$ is a fixed initial condition.
These variables can be discrete or continuous in our general setup; in the continuous case, summations appearing below should be replaced by integrals.

The dynamics proceeds as follows.
At each time step:
\begin{enumerate}
  \item $h_t$ is updated deterministically:
    $h_t = \Phi_t(y_1,  y_2,  \dots,  y_t)$ for $t=1,2,\dots,T$, and $h_0 = \Phi_0 ( \ )$ is set to a constant.
  \item $y_{t+1}$ is drawn from the conditional distribution $p_t(y_{t+1} \mid h_{t})$, also called the \emph{emission kernel}, for $t=1,\dots,T-1$, and $p_0 (y_1 \mid h_0 ) \equiv p(y_1)$ is the initial distribution. 
\end{enumerate}

Here, each $\Phi_t$ is a deterministic map that accepts a variable-length
sequence of observations and returns a latent state.
The subscript $t$ indexes parameters of the map (e.g., its learnable parameters),
but \emph{not} the length of the input sequence;
indeed, in the backward process (Section~\ref{sec:backward}),
$\Phi_t$ will be applied to an input whose length differs from~$t$.
For Transformers, $\Phi_t$ is independent of~$t$
(i.e., the same attention mechanism with shared parameters
is applied at every step), and can process input sequences of
arbitrary length within the context window.

The marginal process $y_t$ is in general non-Markovian,
since the latent state $h_t$ accumulates information from past observations.
The joint process $(h_t, y_t)$ is  generally not Markovian either,
 because $h_t = \Phi_t(y_1, \dots, y_t)$
cannot in general be determined from $(h_{t-1}, y_t)$ alone.

We now rewrite the update of the latent state $h_{t}$ as
\begin{equation}\label{eq:h-forward}
  h_{t} = f_t^{\to}(y_{1:t}),
\end{equation}
where  $y_{1:t} \equiv (y_1, y_2, \dots, y_{t})$ and
\begin{equation}\label{eq:f-forward-def}
  f_t^{\to}(y_{1:t})  \equiv  \Phi_{t}(y_1,  \dots,  y_{t}).
\end{equation}
The reason why we introduce the separate notation $f_t^{\to}$ is that $\Phi_{t}$ will be used for the definition of the backward process as well.

\begin{figure}[htbp]
  \centering
  \includegraphics[width=1.0\linewidth]{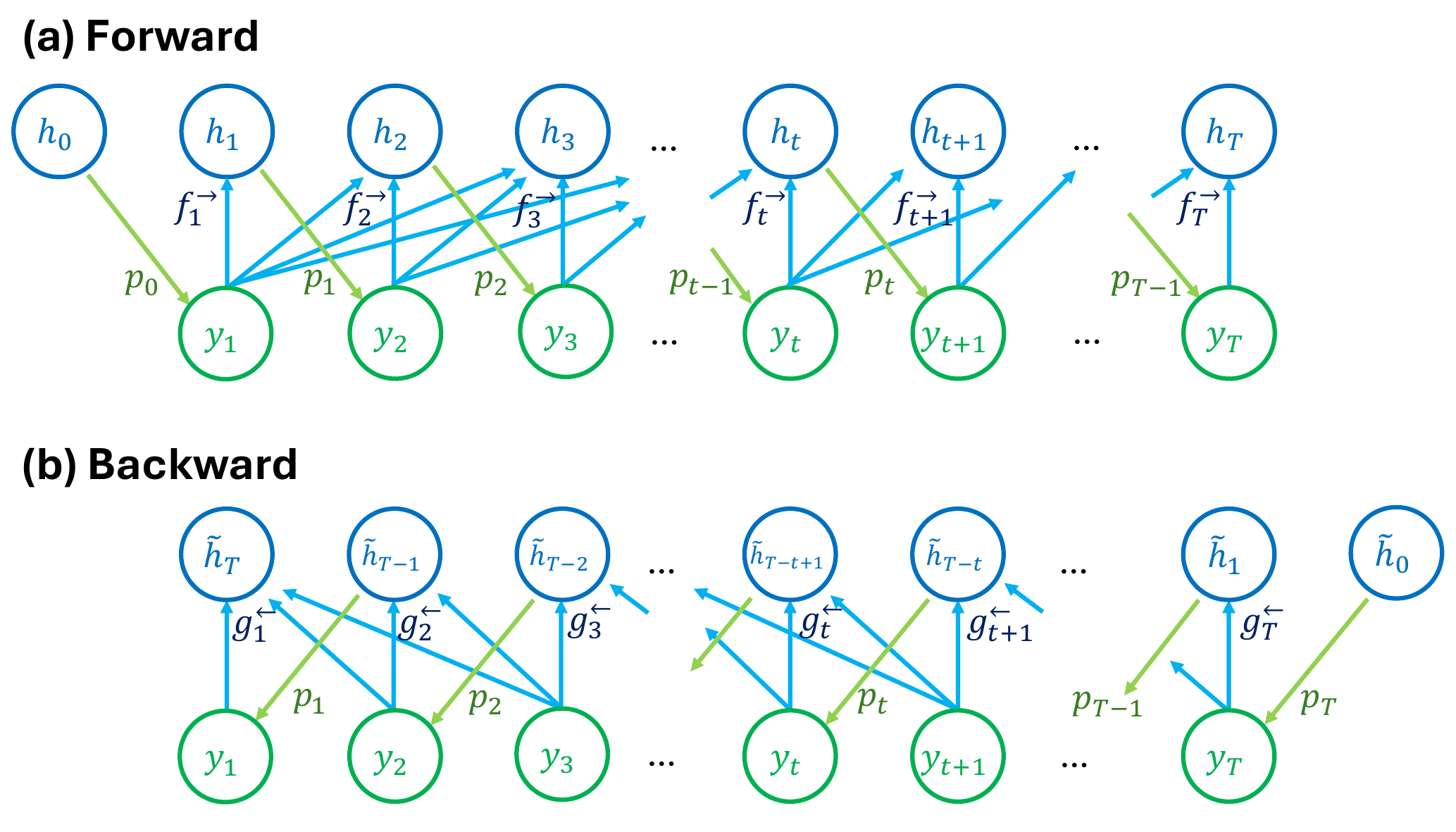}
  \caption{Schematic of the general causal structure of our setup for the general (non-recursive) case. (a) the forward process~\eqref{eq:P-forward}, and (b) the backward process \eqref{eq:P-backward}. The blue arrows indicate deterministic functions, while the green arrows indicate stochastic influences. This figure illustrates the particular realization $\tilde y_s = y_{T-s+1}$; even in this case, $\tilde h_s \neq h_{T-s+1}$ in general.}
  \label{fig:causal}
\end{figure}

The forward process generates the sequence $y_{1:T} = (y_1, y_2, \dots, y_{T})$
by iterating the update--emission rule above.
Its path probability is
\begin{equation}\label{eq:P-forward}
  P_{\to}(y_{1:T})
  = \prod_{t=0}^{T-1} p_t \bigl(y_{t+1} \mid f_t^{\to}(y_{1:t})\bigr).
\end{equation}
See Figure~\ref{fig:causal} (a) for a graphical representation.

We note that if one simply defined $h_t = (y_1, y_2, \dots, y_t)$ (i.e., $\Phi_t$ were set to the identity map on the entire history),
the resulting model could represent an arbitrary non-Markovian process on $y_t$.
However,  the important restriction of our architecture lies in treating
$h_t$ as having a  \emph{fixed finite state space} (or, for continuous variables, a \emph{fixed finite dimensionality})  independent of $t$:
$h_t$ must encode the growing history
$y_{1:t}$ into a representation of bounded size.
Note that this constraint concerns the size of $h_t$ alone;
the input sequence to $\Phi_t$ may be of arbitrary length.

Because the emission kernel $p_t(y_{t+1}\mid h_t)$ depends on the
history only through the deterministic latent state~$h_t$, each factor in
the forward (and backward) path probabilities~\eqref{eq:P-forward}
(and~\eqref{eq:P-backward} below) is directly evaluable for any given
trajectory, and the entropy production $\mathcal{S}_y$ can be
estimated by Monte Carlo sampling without an exponential cost from enumerating histories,
 as shown in Section~\ref{sec:sampling}.

Moreover, the latent state $h_t$ is not merely a convenient computational device
but plays the role of a \emph{sufficient statistic} of the past
$y_{1:t}$ for predicting the next observation $y_{t+1}$.
Indeed, the conditional distribution of $y_{t+1}$ given the entire
history $y_{1:t}$ factorizes through $h_t$ by construction:
$P_{\to}(y_{t+1} \mid y_{1:t}) = p_t(y_{t+1} \mid h_t)$,
which is precisely the defining property of a sufficient statistic.
Thus, the bounded-size compression from $y_{1:t}$ to $h_t$
incurs no loss of predictive information in the forward direction,
regardless of how long the history is.
The interplay between sufficient statistics and entropy production
in Markovian systems has been studied
in~\cite{MatsumotoSagawa2018}.

\subsection{Recursive case}\label{sec:recursive-special-case}

\begin{figure}[htbp]
  \centering
  \includegraphics[width=1.0\linewidth]{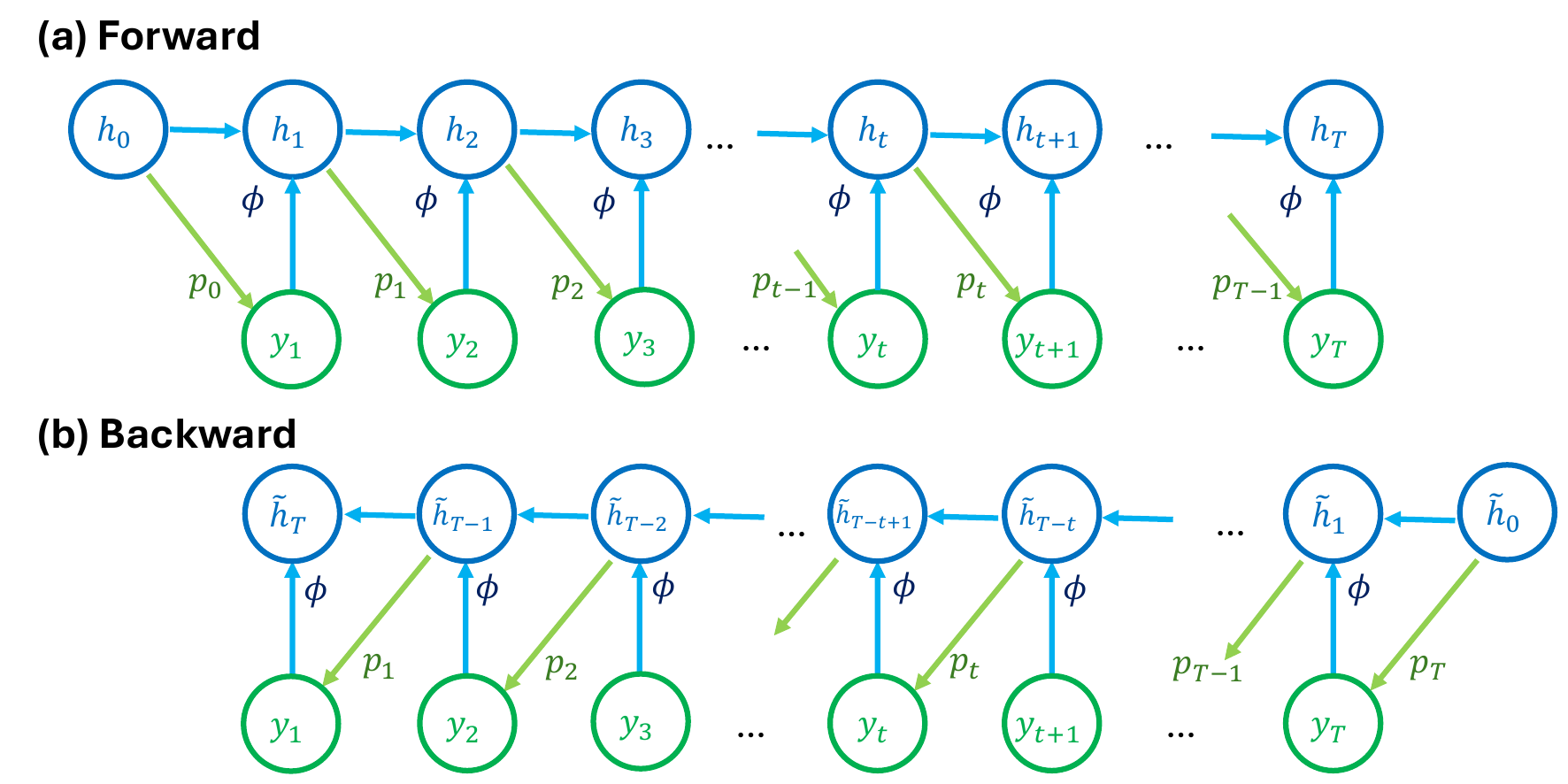}
  \caption{Schematic of the causal structure for the recursive case.  (a) the forward process~\eqref{eq:P-forward} with \eqref{r_h_f}, and (b) the backward process \eqref{eq:P-backward} with \eqref{re_backward}. The blue arrows indicate deterministic functions, while the green arrows indicate stochastic influences. By following the arrows, one can see that this recursive diagram is a special case of the general diagram (Figure~\ref{fig:causal}).  This figure illustrates the particular realization $\tilde y_s = y_{T-s+1}$; even in this case, $\tilde h_s \neq h_{T-s+1}$ in general. }
  \label{fig:causal2_new}
\end{figure}

An important special case is that $h_t$ is determined only by $(h_{t-1}, y_t)$ through a deterministic function $\phi$ as
\begin{equation}\label{r_h_f}
h_t = \phi (h_{t-1}, y_t).
\end{equation}
Here, we assume that $\phi$ is time-independent, but the emission kernel $p_t$ may still depend on time.
Correspondingly,  $\Phi$ factors as
\begin{equation}\label{eq:recursive-Phi}
  \Phi(y_1,  \dots,  y_t)
  = \phi \bigl(\Phi (y_1,  \dots,  y_{t-1}),  y_t\bigr).
\end{equation}
The graphical representation of the dependency structure is shown in Figure~\ref{fig:causal2_new} (a).

In this case, the joint process $(h_t, y_t)$ is Markovian (see also Appendix~\ref{sec:markovian-embedding}).
Moreover, the marginal dynamics of $h_t$ alone is Markovian if $y_t$ is marginalized, as is the case for the predicted state $\hat{x}_{t+1|t}$ in the Kalman filter.
Even so, the marginal dynamics of $y_t$ is non-Markovian in general.

Most of the results below hold in the general (non-recursive) setting;
when additional structure specific to the recursive case is relevant,
it will be noted explicitly.

%----------------------------------------------------------------------
\subsection{Examples}\label{sec:examples}
%----------------------------------------------------------------------

We show that several well-known
architectures can indeed be regarded as special cases of our general setup.
In all cases below, implementation details
(e.g., layer normalization, multi-head decomposition, residual connections for Transformers)
are omitted, and only a simplified description is given.
The correspondence between the general framework and each example
is summarized in Table~\ref{tab:examples}.

\begin{table*}[t]
\centering
\small
\renewcommand{\arraystretch}{1.6}
\caption{Correspondence between our general framework and representative
 architectures.
In all cases $y_t$ denotes the observed variable (token or measurement).
The last column indicates whether the recursive special case holds ($h_t = \phi (h_{t-1}, y_t)$).
}
\label{tab:examples}
\smallskip
\begin{tabular}{l p{2.2cm} p{4.6cm} p{3.8cm} c}
\hline
\textbf{Model}
  & $h_{t}$
  & $\Phi_{t}$ or $\phi$
  & $p_t(y_{t+1} \mid h_{t})$
  & \textbf{Recursive} \\
\hline
Transformer
  & Attention \newline context
  & $\Phi (y_1, \dots, y_t)$\newline [full sequence]
  & $\mathrm{softmax}(W_{\mathrm{out}} h_{t})_{y_{t+1}}$
  & $\times$ \\
RNN
  & RNN \newline latent state
  & $\phi(h,y)=$\newline $\tanh (W_h h + W_y y + b)$
  & $\mathrm{softmax}(W_{\mathrm{out}} h_{t} + b_{\mathrm{out}})_{y_{t+1}}$
  & $\checkmark$ \\
Kalman
  & $\hat{x}_{t+1|t}$
  & $\phi (h,y)=$\newline $A[(I - K C)h + K y]$
  & $\mathcal{N}(C h_{t},  S)$
  & $\checkmark$ \\
SSM
  & SSM state
  & $\phi (h,y)=$\newline $A h + B y$
  & $\mathrm{softmax}(W_{\mathrm{out}} C h_{t})_{y_{t+1}}$
  & $\checkmark$ \\
Mamba
  & $h_t' \equiv (h_t,\, y_t)$
  & $\phi'(h',y)=$\newline
    $\bigl(A(y)\, h + B(y)\, y,\; y\bigr)$
  & $p(y_{t+1} \mid h_{t}') =$ \newline $\mathrm{softmax}\bigl(W_{\mathrm{out}}\, C(y_t)\, h_{t}\bigr)_{y_{t+1}}$
  & $\checkmark$ \\
\hline
\end{tabular}
\end{table*}

\noindent\textbf{Transformers.}
In a simplified autoregressive Transformer,
the observed variable $y_t$ is a discrete token taking values
in a finite vocabulary $\mathcal{Y}$.
Each token is first mapped to a continuous vector
by an embedding matrix $E \in \mathbb{R}^{d \times |\mathcal{Y}|}$,
and causal self-attention then computes a context vector
$h_{t} = \Phi(y_1,\dots,y_{t}) \in \mathbb{R}^{d}$
from the full past sequence.
(Concretely, the token $y_t \in \mathcal{Y}$ is first encoded
as a one-hot vector $e_{y_t} \in \{0,1\}^{|\mathcal{Y}|}$,
and the embedding is the matrix--vector product $E e_{y_t}$;
the variable $y_t$ itself remains a discrete label
throughout the framework.)
Thus $y_t$ is discrete while $h_t$ is continuous;
the embedding step is absorbed into the deterministic map~$\Phi$.
The next token is drawn from the emission kernel
$p(y_{t+1} \mid h_{t})
  = \mathrm{softmax}(W_{\mathrm{out}}\, h_{t})_{y_{t+1}}$,
where the subscript $y_{t+1}$ selects
the component of the softmax vector
corresponding to the token $y_{t+1} \in \mathcal{Y}$.

Since $\Phi$ accesses the entire history and cannot in general
be reduced to a function of $(h_{t-1}, y_{t})$,
this is an instance of the general (non-recursive) case:
$h_t$ corresponds to the general map $\Phi$,
and the joint process $(h_t, y_t)$ is not Markovian.
Note that the model parameters for $\Phi_t$ and $p_t$ are both time-independent in Transformers.
(While the positional encoding added to each token embedding
depends on the token's position within the input sequence to $\Phi$,
this position is the argument index of the function $\Phi$ and not its external time label $t$.)

We remark that, in a multi-layer Transformer, the key--value pairs cached at each layer up to step~$t$
(the KV-cache) constitute an alternative representation of the past.
If one redefines the latent state as the full KV-cache, denoting it by $h_t^{\mathrm{KV}}$,
then the update at step~$t$ computes new key--value pairs at each layer from
$h_{t-1}^{\mathrm{KV}}$ and~$y_t$, and the emission kernel
$p(y_{t+1} \mid h_t^{\mathrm{KV}})$ is read off from the final-layer output.
The resulting update $h_t^{\mathrm{KV}} = \varphi(h_{t-1}^{\mathrm{KV}}, y_t)$ thus formally
satisfies the recursive structure.
However, $h_t^{\mathrm{KV}}$ consists of $t$ key--value pairs per layer per head, so the
dimensionality of the state space grows linearly with~$t$,
violating the fixed-size requirement on~$h_t$ for large $t$.

\noindent\textbf{RNN.}
An Elman-type RNN updates its latent state by a two-argument recurrence
$h_t = \phi(h_{t-1}, y_t) = \tanh (W_h h_{t-1} + W_y y_t + b)$ (a nonlinear activation function)
and predicts via
$p(y_{t+1} \mid h_{t}) = \mathrm{softmax}(W_{\mathrm{out}}  h_{t} + b_{\mathrm{out}})_{y_{t+1}}$.
This is the recursive special case
(Section~\ref{sec:recursive-special-case}).
As in the Transformer case, $y_t$ is a discrete token
and $h_t \in \mathbb{R}^d$ is a continuous latent vector;
the product $W_y\, y_t$ implicitly performs embedding
 via the one-hot encoding of $y_t$.

\noindent\textbf{Steady-state Kalman filter.}
Consider a linear Gaussian system
$x_{t+1} = A x_t + w_t$,
$y_t = C x_t + v_t$
with $w_t \sim \mathcal{N}(0,Q)$,
$v_t \sim \mathcal{N}(0,R)$ independent.
The true state $x_t$ has no counterpart
in our framework;
we interpret the Kalman filter as a generative model that reproduces the trajectory distribution of $y_t$.

In the innovation representation of the Kalman filter, we set
$h_{t-1}$ as $\hat{x}_{t|t-1}$
(the one-step-ahead prediction).
We also assume that the Kalman filter has reached steady state.
Then
$p(y_t \mid h_{t-1}) = \mathcal{N}(C h_{t-1},  S)$
with $S = C P C^\top + R$, where $P$ is the steady-state prediction error covariance.
The recursive map is the linear update
$\phi \colon (h, y) \mapsto A[(I - K C) h + K  y]$,
where $K = P C^\top S^{-1}$.
This is the recursive special case with linear activation and Gaussian output.
In contrast to the Transformer and RNN cases,
both $y_t$ and $h_t$ are continuous variables here.
See Section~\ref{sec:linear-gaussian-example} for details.

\noindent\textbf{SSM.}
By ``SSM'' we mean a discrete-time linear state space model layer
(e.g.~\cite{GuGoel2022,SmithWarringtonLinderman2023})
whose parameters $A$, $B$, $C$ are time-independent.
The latent state is updated by
$h_t = A h_{t-1} + B y_t$,
and the next-token distribution is obtained from $Ch_t$
via an output projection and softmax.
This is the recursive special case with
$\phi \colon (h, y) \mapsto A h + B y$ (where $y$ stands for its one-hot encoding $e_y$, as in the Transformer case),
which is linear in both arguments.
Since $C$ is independent of the input $y_t$, the emission kernel
$p(y_{t+1} \mid h_t)
= \mathrm{softmax}(W_{\mathrm{out}} C h_t)_{y_{t+1}}$
depends on $h_t$ alone.
As in the Transformer and RNN cases,
$y_t$ is a discrete token while $h_t \in \mathbb{R}^d$
is continuous.

\noindent\textbf{Mamba (selective SSM).}
Mamba~\cite{GuDao2023} introduces input-dependent (``selective'')
parameters: $A(y_t)$,
$B(y_t)$, and $C(y_t)$ are all functions
of the current input $y_t$.
While the state update
$h_t = A(y_t)\, h_{t-1} + B(y_t)\, y_t$
remains a deterministic function of $(h_{t-1}, y_t)$,
the output projection involves $C(y_t)\, h_t$,
so that the emission kernel depends on $y_t$ as well as $h_t$.
Since $y_t$ cannot in general be recovered from $h_t$ alone,
the SSM state $h_t$ by itself is not a sufficient statistic of
the past for predicting $y_{t+1}$.

To accommodate Mamba within the recursive framework,
it suffices to augment the latent state as
$h'_t \equiv (h_t,\, y_t)$.
The augmented update is $h'_t = \phi'(h'_{t-1}, y_t)
       = \bigl(A(y_t)\, h_{t-1} + B(y_t)\, y_t,\; y_t\bigr)$.
This is a deterministic function of $(h'_{t-1}, y_t)$ and preserves
the recursive structure.
As in the SSM case,
$y_t$ is discrete and $h_t$ is continuous,
so the augmented state $h'_t = (h_t,\, y_t)$
comprises both continuous and discrete components.

%======================================================================
\section{Backward process and entropy production}
\label{sec:backward_ep}
%======================================================================

In this section, we construct the backward process by reusing the same architectural components in reversed temporal order and define the entropy production as the KL divergence between the forward and backward path measures.

\subsection{Backward process}\label{sec:backward}

The backward process is defined   by reusing the same
emission kernels $p_t$ and deterministic maps $\Phi_t$ in the backward direction to produce a sequence in the reversed temporal order.
Concretely, we fix an initial latent state $\tilde{h}_0$ for the backward process
and generate a sequence $(\tilde{y}_1, \tilde{y}_2, \dots, \tilde{y}_{T})$
according to:
\begin{enumerate}
  \item $\tilde h_s$ is updated deterministically:
    $\tilde h_s = \tilde \Phi_{s} (\tilde y_1,  \tilde y_2,  \dots,   \tilde y_s)$ for $s=1,2,\dots,T$, and $\tilde h_0 = \tilde \Phi_0 ( \ )$ is set to a constant (initial condition).
  \item $\tilde y_{s+1}$ is drawn from the conditional distribution
    $\tilde p_{s}(\tilde y_{s+1} \mid \tilde h_{s})$ for $s=1,\dots,T-1$, and $\tilde p_0 (\tilde y_1 \mid \tilde h_0 ) \equiv \tilde p ( \tilde y_1)$ is the initial distribution of the backward process. 
\end{enumerate}

The backward path probability is then given by 
\begin{equation}\label{eq:P-backward0}
  P_{\leftarrow}(\tilde y_{1:T})
  = \prod_{s=0}^{T-1} \tilde p_{s} \bigl(\tilde y_{s+1} \mid \tilde \Phi_s (\tilde y_1,  \tilde y_2,  \dots,   \tilde y_s) \bigr).
\end{equation}

We now relate these notations to those in the forward process.
We interpret this backward process as generating the forward-time sequence
in reversed order.
That is, we  identify 
\begin{equation}
\tilde y_s = y_{T-s+1}.
\end{equation}
We also set
\begin{equation}
\tilde p_s = p_{T-s} \ (s=1, \dots, T-1), \tilde \Phi_s = \Phi_{T-s+1} \ (s=1, \dots, T),
\end{equation}
which is consistent with the standard notion of backward protocols in stochastic thermodynamics (as explicitly shown later).
We note that we assume that $y$ and $h$ do not involve any parity-odd quantity that changes sign under time-reversal (such as the momentum of the underdamped Langevin equation).

As mentioned above, the initial distribution of the backward process is given by
\begin{equation}\label{eq:backward_initial0}
\tilde p ( \tilde y_1) \equiv \tilde p_0 (\tilde y_1 \mid \tilde h_0 ) ,
\end{equation}
where $\tilde h_0$ is a fixed constant. In our framework, it is natural to assume that $\tilde p_0$ is a known function chosen independently of $p(y_T)$ in the forward process.
On the other hand, another natural choice in stochastic thermodynamics is
\begin{equation}\label{eq:assume_final}
\tilde p ( \tilde y_1) = p(y_T),
\end{equation}
 which means that the initial distribution of the backward process is identical to the final distribution of the forward process. 
 We do \emph{not}  assume \eqref{eq:assume_final} in this paper unless otherwise stated, because it is operationally intractable in our framework in general.
 We emphasize that, also in the standard formulation of stochastic thermodynamics,  \eqref{eq:assume_final}  is not always assumed (e.g.,~\cite{KawaiParrondoBroeck2007}), and even in such cases the entropy production still has a  clear physical meaning, especially when the initial distribution of the backward process is chosen to be the Gibbs distribution with respect to a fixed Hamiltonian.
This point will be discussed in Section~\ref{subsec:autoregressive-tractable} in more detail, and a concrete example will be illustrated in Section~\ref{sec:linear-gaussian-example}.

Since each $\Phi_t$ accepts variable-length input,
the map $\Phi_{T-s+1}$ can be applied to the $s$-element backward sequence
$(\tilde{y}_1, \dots, \tilde{y}_s)$, using the parameters
(e.g.\ attention weights) indexed by $T-s+1$.
The backward process thus ``runs the same machinery in reverse'':
the same operators $\Phi_t$ and kernels $p_t$ are reused,
but they are invoked in the order $t = T, T-1, \dots, 1, 0$,
and their input is the backward sequence accumulated so far.
Note that $\Phi_{T+1} ( \ ) = \tilde \Phi_0 ( \ )$ has no counterpart in the forward process, but it just gives a  chosen constant $\tilde h_0$.

Then,
\begin{equation}\label{eq:P-backward-s}
  P_{\leftarrow}( y_{T:1})
  = \prod_{s=0}^{T-1} p_{T-s} \bigl( y_{T-s} \mid  \Phi_{T-s+1} ( y_T,   y_{T-1},  \dots,    y_{T-s+1}) \bigr).
\end{equation}
Here, $p_T$ for $s=0$ is introduced solely to denote the initial distribution of the backward process: $p_T (y_T) \equiv \tilde p ( y_T) \equiv \tilde p_0 ( y_T \mid \tilde h_0 )$.
We re-index by $t = T - s $ (i.e., $s = T - t $) to express this in
forward-time indices, and introduce the notation 
\begin{equation}
  g_{t+1}^{\leftarrow}( y_{T:t+1})  \equiv  \Phi_{t+1}( y_T,   y_{T-1},  \dots,    y_{t+1})  = \tilde{h}_{T-t}.
\end{equation}
We finally obtain
\begin{equation}\label{eq:P-backward}
  P_{\leftarrow}(y_{T:1})
  = \prod_{t=1}^{T} p_{t} \bigl(y_{t} \mid g_{t+1}^{\leftarrow}( y_{T:t+1}) \bigr).
\end{equation}
See Figure~\ref{fig:causal} (b) for a graphical representation.

We emphasize that we cannot assume $\tilde h_s = h_{T-s+1}$ even for the event that the time-reversed sequence $\tilde y_s = y_{T-s+1}$ is realized.
This is because the reverse run of $y_t$ does not necessarily produce the reverse run of $h_t$ by applying the given deterministic function $\Phi_t$ in the reverse way, even for the recursive case.

It is also important to distinguish the backward process introduced here from Bayesian retrodiction $P_{\to}(y_t \mid y_{t+1:T})$ (see Section~\ref{sec:retro}); rather, in direct analogy with Crooks-type stochastic thermodynamics, the backward process is obtained by reversing the protocol  implemented by  $p_t$ and $\Phi_t$.

\subsection{Entropy production}

We now define the entropy production of the observed sequence as the KL divergence between the forward and backward path measures:
\begin{equation}\label{eq:Sigma-def}
\begin{split}
  \mathcal{S}_y
  &\equiv
 D_{\mathrm{KL}}  
   \bigl(P_{\to}(y_{1:T})  \big\|  P_{\leftarrow}(y_{T:1}) \bigr) \\
  &=  \mathbb{E}_{P_{\to}} \left[
    \ln \frac{P_{\to}(y_{1:T})}{P_{\leftarrow}(y_{T:1})}
  \right] \\
  &\geq  0.
   \end{split}
\end{equation}
Non-negativity follows from the general property of KL divergence.
This can be equivalently represented as
\begin{equation}
\begin{split}
  \mathcal{S}_y
  &= \mathbb{E}_{P_{\to}} \left[
    \ln \frac{ \prod_{t=0}^{T-1} p_t \bigl(y_{t+1} \mid f_t^{\to}(y_{1:t}) \bigr)}{\prod_{t=1}^{T} p_{t} \bigl(y_{t} \mid g_{t+1}^{\leftarrow}( y_{T:t+1}) \bigr)} 
  \right]   \\
  &=  \mathbb{E}_{P_{\to}} \left[ - \ln \tilde p (y_T) + \ln p (y_1)   \right] \\
  &{} \quad  + \sum_{t=1}^{T-1}\mathbb{E}_{P_{\to}}
  \left[
    \ln \frac{ p_t \bigl(y_{t+1} \mid f_t^{\to}(y_{1:t}) \bigr)}{ p_{t} \bigl(y_{t} \mid g_{t+1}^{\leftarrow}( y_{T:t+1}) \bigr)}  \right]. \label{EP_property}
\end{split}
\end{equation}
If we assume \eqref{eq:assume_final}, the first term becomes
\begin{equation}
\begin{split}
 &\mathbb{E}_{P_{\to}} \left[ - \ln p (y_T) + \ln p (y_1)   \right] \\
 &{} \quad = -\sum_{y_T} p (y_T) \ln p (y_T) + \sum_{y_1} p (y_1) \ln p (y_1),
 \end{split}
\end{equation}
which is the change in the Shannon entropy of the single-time marginal distribution in the forward process.

Correspondingly, the stochastic entropy production is defined as
\begin{equation}
\sigma(y_{1:T})
  \equiv
  \ln \frac{P_{\to}(y_{1:T})}{P_{\leftarrow}(y_{T:1})},
\end{equation}
which gives 
\begin{equation}
  \mathcal{S}_y= \mathbb{E}_{P_{\to}} [ \sigma(y_{1:T}) ].
\end{equation}
The integral fluctuation theorem~\cite{Seifert2005,Jarzynski1997,Crooks1999} is automatically satisfied:
\begin{equation}
\mathbb{E}_{P_{\to}} [ e^{- \sigma(y_{1:T})} ] = 1.
\end{equation}

Here, rigorously speaking, the integral fluctuation theorem requires that $P_{\to}(y_{1:T}) > 0$ whenever $P_{\leftarrow}(y_{T:1}) > 0$. This condition is satisfied in all examples of Table~\ref{tab:examples}: for discrete-token models (Transformers, RNNs, SSMs, Mamba), the softmax emission kernel assigns strictly positive probability to every token, while for the linear Gaussian case the forward and backward path measures share the same support by construction.

We note that $\mathcal{S}_y$ depends not only on the forward path measure
$P_{\to}(y_{1:T})$ but also on the specific decomposition into emission
kernels $p_t$ and deterministic maps $\Phi_t$, since the backward
process~\eqref{eq:P-backward} reuses these architectural components in
reversed temporal order.
This parallels a situation in stochastic thermodynamics for
non-Markovian processes: the entropy production based on Crooks-type
protocol reversal depends on the specific underlying dynamical
model, such as the memory kernel and bath structure in generalized
Langevin systems~\cite{SpeckSeifert2007,OhkumaOhta2007}.
In the Markovian limit, the backward transition kernel
$P_{\leftarrow}(y_{t} \mid y_{t+1})$ is naturally
determined by the forward process, consistent with the reduction to the standard Crooks
formulation~\cite{Crooks1999} as shown below.

\subsection{Recursive case}
\label{sec:recursive_b}

In the recursive special case of Section~\ref{sec:recursive-special-case},
we choose the backward model so that its
latent state is updated recursively by
\begin{equation}\label{re_backward}
\tilde h_s = \phi(\tilde h_{s-1}, \tilde y_s).
\end{equation}
The graphical representation of the dependency structure is shown in Figure~\ref{fig:causal2_new} (b).
We also assume that $h_0 = \tilde h_0$.

As mentioned before, the reverse run of $y_t$ does not necessarily produce the reverse run of $h_t$ by applying the given deterministic function $\phi$.  To explicitly see this, the following small example suffices: let $T=2$, $\phi (h,y) \equiv 2h +3y$, and $h_0 = \tilde h_0 = 0$.  Then, $h_1 = 3y_1$, $h_2 = \phi (h_1, y_2) =  \phi (3y_1, y_2) = 6 y_1 +3 y_2$.  In the reverse run, $\tilde h_1 = 3 y_2$, $\tilde h_2 = \phi (\tilde h_1, y_1) = \phi (3y_2, y_1) = 3y_1 + 6y_2 $.  Obviously $h_1 \neq \tilde h_2$ and $h_2 \neq \tilde h_1$.

This mismatch has an important implication when one considers the Markovian embedding $\mathbf{x}_t \equiv (h_t, y_t)$, which is available in the recursive case (see Appendix~\ref{sec:markovian-embedding} for details).
From the above consideration, the forward and backward transition kernels of $\mathbf{x}_t$ have disjoint support in general, so that the entropy production $\mathcal{S}_{\mathbf{x}}$ defined at the $\mathbf{x}$-level may diverge and would not be informative.
Even so, the entropy production of $y$ itself, $\mathcal{S}_y$, stays finite, as explicitly shown in Section~\ref{sec:linear-gaussian-example} for a concrete example.
This provides an additional motivation for our approach, which defines the entropy production $\mathcal{S}_y$ solely through the path probabilities of the observed sequence $y_{1:T}$, without relying on a Markovian embedding.

\noindent\textbf{Markovian case.}
As a further specific subclass of the recursive class, we remark on the case where $y_t$ itself is Markovian.
Indeed, the standard formulation of stochastic thermodynamics of Markovian dynamics can be regarded as a special case of our formulation, where we can take $\Phi (y_1, \dots, y_t) = y_t$ so that $h_{t} = y_t$ for all $t$.
Then automatically $ g_{t+1}^{\leftarrow}( y_{T:t+1}) =  \Phi( y_T,   y_{T-1},  \dots,    y_{t+1}) = y_{t+1}$. 
Equivalently, $\phi (h,y) = y$ gives $h_t = y_t$ and $\tilde h_s = \tilde y_s$. 
Therefore, the second term of \eqref{EP_property} reduces to
\begin{equation}
 \sum_{t=1}^{T-1}\mathbb{E}_{P_{\to}}
  \left[
    \ln \frac{ p_t \bigl(y_{t+1} \mid y_t)}{ p_{t} \bigl(y_{t} \mid y_{t+1})}  \right],
\end{equation}
which is exactly equivalent to the standard definition in Markovian stochastic thermodynamics \textit{\`{a} la} Crooks~\cite{Crooks1999}.
Here, it is important that the forward transition $y_t \to y_{t+1}$ and the backward transition $y_{t+1} \to y_t$ are induced by the same kernel $p_t$ with index $t$.
If we further assume the local detailed balance condition, this term is regarded as $-\beta Q$ with $\beta$ being the inverse temperature and $Q$ being the heat absorbed from a thermal reservoir.
We emphasize that we assume neither Markovianity nor detailed balance  in our general framework.

\subsection{Discussion}

Let us clarify the relation of our definition of the entropy production~\eqref{eq:Sigma-def} to existing notions.
A line of work directly related to the present approach quantifies irreversibility on the basis of the KL divergence between the forward and backward processes~\cite{KawaiParrondoBroeck2007}.
Seminal works~\cite{RoldanParrondo2010,RoldanParrondo2012} adopted this approach for the observed non-Markovian process; see also the stationary Gaussian analysis of \cite{SearaMachtaMurrell2021}. Closely related observed-path constructions arise from marginal or coarse-grained path measures, including lower bounds from coarse-grained trajectories \cite{GomezMarinParrondoVandenBroeck2008,KawaguchiNakayama2013,KahlenEhrich2018,CrooksStill2019}. Another approach treats non-Markovian physical dynamics by explicitly assuming the presence of thermal reservoirs, especially generalized Langevin systems with colored baths~\cite{SpeckSeifert2007,OhkumaOhta2007}.

By contrast, in our deterministic-memory autoregressive setting, each factor of the forward/backward path-probability ratio is supplied directly by the model's emission kernel together with the deterministic memory update.  As a consequence, $\mathcal{S}_y$ is directly computable from the model itself without empirical estimation of long-history conditionals, and without introducing an auxiliary stochastic hidden-state dynamics.

Because our framework does not presuppose any thermal reservoir, we impose no local detailed balance condition.
In particular, when the framework is applied to an LLM, $\mathcal{S}_y$ does not directly quantify the physical energy cost or heat dissipation of the hardware that executes the model.
If one regards the full physical state of the computing device as a Markovian process $x_t$ that generates the token sequence $y_t$ (see  Appendix~\ref{sec:markovian-embedding} and assumptions there), the coarse-graining inequality $\mathcal{S}_x \geq \mathcal{S}_y$ implies that $\mathcal{S}_y$ may serve as a lower bound on the physical entropy production, but the two quantities are in general distinct.
For a concrete physical implementation, the actual dissipation depends on hardware-specific details and is therefore not determined by $\mathcal{S}_y$ alone.  Quantifying the tightness and operational significance of this bound for specific hardware architectures remains an open problem. Constructing a simplified physical model that explicitly implements the autoregressive update and comparing  $\mathcal{S}_x$ and $\mathcal{S}_y$ would be a worthwhile direction for future work.

In all architectures listed in Table~\ref{tab:examples}, the only externally observable variable
is $y_t$.
The latent state $h_t = \Phi_t(y_1,\dots,y_t)$ is a deterministic
function of the observed history and carries no independent stochastic
degrees of freedom.
On the other hand, behind the observations $y_t$ there may exist a true environmental
state $x_t$ whose dynamics generates $y_t$.
The Kalman filter example (Section~\ref{sec:linear-gaussian-example})
makes this explicit: $x_t$ evolves as
$x_{t+1} = A x_t + w_t$ and produces observations
$y_t = C x_t + v_t$, but $x_t$ has no counterpart in the
framework.
In the case of an LLM, there are at least two fundamentally different candidates for such a state $x_t$: the physical internal state of the hardware executing the model, and the state of the real-world process that the text describes.

Meanwhile, the asymmetry between forward and backward
modeling of natural language has been studied from
machine-learning perspectives.
Ref.~\cite{PapadopoulosWengerHongler2024}
quantified  the irreversibility of language models by separately training the backward  model on the time-reversed dataset of natural language,
which is distinct from our definition of the backward process for which the same model as the forward process is used.
Ref.~\cite{YuEtAl2025} computed the per-trajectory difference
between forward and backward losses under a single model,
but the model itself is trained on both
directions.
Note that neither work formulated a connection to stochastic
thermodynamics.

We also remark on terminology.
Ref.~\cite{MoslonkaRandrianarivGarnier2025} introduced a quantity called the entropy production rate for LLMs, defined as the average Shannon entropy of per-token predictive distributions; this involves neither time reversal nor a latent-state architecture, and is conceptually distinct from our entropy production $\mathcal{S}_y$~\eqref{eq:Sigma-def}.

%======================================================================
\section{Estimation of entropy production}
\label{sec:sampling}
%======================================================================

The entropy production $\mathcal{S}_y$ defined in \eqref{eq:Sigma-def}
involves the log-ratio of the forward and backward path probabilities
summed over all time steps.
A central question for practical applications --- particularly for LLMs
and other large-scale autoregressive models --- is whether $\mathcal{S}_y$
 can be estimated from a finite number of sampled
trajectories.
In this section, we show that the structure of the
autoregressive framework (Section~\ref{sec:setup}) makes $\mathcal{S}_y$
itself efficiently computable by standard Monte Carlo sampling.
This sampling method will be applied to GPT-2 in Section~\ref{subsec:gpt2-generated} as a proof-of-concept demonstration with a pre-trained language model, and will be demonstrated in Section~\ref{subsec:numerical-verification} for the linear Gaussian case. 
We further show that temporal coarse-graining, in which the backward
pass reverses the order of blocks rather than individual tokens,
can be evaluated at the same per-trajectory cost
(Section~\ref{sec:implications-cg}).

%----------------------------------------------------------------------
\subsection{Sampling cost for general non-Markovian processes}
\label{subsec:general-intractable}
%----------------------------------------------------------------------

To clarify why the present framework is special, it is instructive
to first consider the generic situation.
Suppose one observes a non-Markovian stochastic process $y_t$
from an unknown source (e.g., a physical experiment) and wishes to
estimate the entropy production of the form
$\mathbb{E}_{P_{\to}}[\sum_t \ln P_{\to}(y_{t+1} \mid y_{1:t}) - \ln P_{\leftarrow}(y_t \mid
  y_{T:t+1})]$.
The conditional probability $P_{\to}(y_{t+1} \mid y_{1:t})$ is then an
unknown function of the entire past history.
Estimating it from data requires observing many trajectories that
share the same prefix $y_{1:t}$; in a continuous 
 or large observation space, the number of such coincidences
is negligible, and the required sample size grows combinatorially with
the length of the conditioning history.
No compression is available unless one makes additional modeling
assumptions.

%----------------------------------------------------------------------
\subsection{Tractability within the autoregressive framework}
\label{subsec:autoregressive-tractable}
%----------------------------------------------------------------------

The framework of Section~\ref{sec:setup} circumvents this difficulty
through the following structural features that make the path
probabilities directly evaluable.

\noindent\textbf{Deterministic latent state.}
The latent state $h_{t} = \Phi_{t}(y_1, \ldots, y_{t})$ is a
deterministic function of the observed history.
Given a single trajectory $y_{1:T}$, every latent state
$h_0, h_1, \ldots, h_{T}$ is uniquely determined without any
stochastic marginalization.
This is in sharp contrast with models with stochastic latent
states, such as hidden Markov models, where
$P(y_{t+1} \mid y_{1:t})
  = \sum_{x_{t+1}} P(y_{t+1} \mid x_{t+1})  P(x_{t+1} \mid y_{1:t})$
involves a sum over latent states $x_{t+1}$ rather than a direct
evaluation from a deterministic state.

\noindent\textbf{Explicit emission kernel.}
The conditional distribution
$p_{t}(y_{t+1} \mid h_{t})$ is an explicit, evaluable function provided
by the model.
For a Transformer-based LLM, this is the softmax output:
\begin{equation}\label{eq:softmax-eval}
  p_t(y_{t+1} \mid h_{t})
  = \frac{\exp \Bigl(\bigl(W_{\mathrm{out}}\, h_{t}\bigr)_{y_{t+1}} / \tau \Bigr)}%
         {\displaystyle\sum_{y} \exp \Bigl( \bigl(W_{\mathrm{out}}\, h_{t}\bigr)_{y} / \tau \Bigr)},
\end{equation}
where $\tau > 0$ is the temperature parameter (usually set to $1$).
No additional sampling or density estimation is needed to
obtain $\ln p_t(y_{t+1} \mid h_{t})$.

\noindent\textbf{Boundary distributions.}
The initial distributions of the forward and backward processes are also explicit. 
In the forward process, the initial distribution is given by $ p(y_1) \equiv p_0 (y_1 \mid h_0 )$, which is a known function of $y_1$ if $p_0$ and $h_0$ are given.
In the backward process, the initial distribution is given by \eqref{eq:backward_initial0}, i.e., $\tilde p ( \tilde y_1) \equiv \tilde p_0 (\tilde y_1 \mid \tilde h_0 )$, which is also a known function of $\tilde y_1$ if $\tilde p_0$ and $\tilde h_0$ are given.
This is precisely the strategy adopted in the linear Gaussian example
of Section~\ref{sec:linear-gaussian-example}, where
$\tilde{p}(\tilde{y}_1) = \mathcal{N}(C \hat{x}_{1|0},\, S)$.
These choices make  the boundary terms directly
evaluable for any sampled trajectory.
On the other hand, if one adopts~\eqref{eq:assume_final}, i.e.,
$\tilde{p}(\tilde{y}_1) = p(y_T)$,
then evaluating this term requires computing the marginal distribution
$p(y_T) = \sum_{y_1,\dots,y_{T-1}} P_{\to}(y_{1:T})$,
which involves a summation (or integration in the continuous case)
over the entire set of earlier observations and is in general
intractable.

%----------------------------------------------------------------------
\subsection{Cost estimates for specific architectures}
\label{subsec:cost-examples}
%----------------------------------------------------------------------
 
Using the product decompositions \eqref{EP_property}, the stochastic entropy production is written as 
\begin{equation}\label{eq:sigma-per-traj}
\begin{split}
  \sigma(y_{1:T})
  =
  &\sum_{t=0}^{T-1}
    \underbrace{%
      \ln p_t \bigl(y_{t+1} \mid f_t^{\to}(y_{1:t}) \bigr)
    }_{\text{forward log-likelihood}}
    \\
    &-
    \sum_{t=1}^T
    \underbrace{%
      \ln p_{t} \bigl(y_{t} \mid g_{t+1}^{\leftarrow}( y_{T:t+1}) \bigr)
    }_{\text{backward log-likelihood}}
    .
    \end{split}
\end{equation}

For a single sampled trajectory $y_{1:T} \sim P_{\to}$,
each term in the sum is obtained as follows.
 
\begin{enumerate}
  \item \textit{Forward pass.} 
    Feed $y_1, y_2, \ldots, y_{T}$ sequentially into the model.
    At each step $t$, the model computes
    $h_{t} =  f_t^{\to}(y_{1:t})$
    and outputs $\ln p_t(y_{t+1} \mid h_{t})$.
  \item \textit{Backward pass (evaluation only).} 
    Feed the reversed sequence $y_{T}, y_{T-1}, \ldots, y_1$
    into the same model (with the same emission kernels $p_t$ and
    maps $\Phi_t$ invoked in reverse temporal order, as specified in
    Section~\ref{sec:backward}).
    At backward step $s$, the model processes
    $y_{T}, \ldots, y_{T-s+1}$, computes
    $\tilde{h}_{s}$, and outputs
    $\ln p_{T-s}(y_{T-s} \mid \tilde{h}_{s})$.
    Note that sampling from the backward process is not necessary to evaluate \eqref{eq:sigma-per-traj}.
\end{enumerate}

Each of the two terms in \eqref{eq:sigma-per-traj} is computationally identical to
a single log-likelihood evaluation: given a sequence
$(y_1, \ldots, y_{T})$, compute the sum
$\sum_t \ln p_t(y_{t+1} \mid h_{t})$ by feeding the tokens
through the model.
This is one of the most basic operations in autoregressive
modeling; it is in fact performed  during training
(where the negative log-likelihood serves as the loss function).
Note that $\sigma(y_{1:T})$ evaluated on a single trajectory $y_{1:T}$, without taking the sample average~\eqref{eq:sigma-MC} below, is itself the stochastic entropy production for that particular realization.
It therefore provides trajectory-level information about the irreversibility of the process.
 
The entropy production is then estimated by Monte Carlo averaging:
\begin{equation}\label{eq:sigma-MC}
  \mathcal{S}_y
   = 
  \mathbb{E}_{P_{\to}}[\sigma]
   \approx 
  \frac{1}{N}\sum_{n=1}^{N} \sigma(y_{1:T}^{(n)}).
\end{equation}
For a target accuracy $\epsilon$, the required number of
sample trajectories scales as
$N = O(\mathrm{Var}(\sigma)/\epsilon^2)$.
Thus the estimator has the standard $N^{-1/2}$ Monte Carlo
convergence, with no additional combinatorial overhead from
enumerating histories.
In general, $\mathrm{Var}(\sigma)$ itself may depend on the
sequence length $T$ and on the statistics of the process.
In Appendix~\ref{app:gpt2-ft}, we numerically demonstrate convergence of the estimator \eqref{eq:sigma-MC} for $T =120$ and $N\simeq 500$.

The total computational cost of the Monte Carlo estimator
\eqref{eq:sigma-MC} is written as
\begin{equation}\label{eq:total-cost}
  C_{\mathrm{total}}
  =
  N  C_1,
\end{equation}
where  $C_1$ is the cost of evaluating $\sigma$
on a single trajectory.
Here and below, ``cost'' refers to the number of floating-point
operations (FLOPs), the standard hardware-independent measure of
arithmetic complexity for numerical computations.
Since $\sigma$ requires one forward and one backward pass, and each
pass is a log-likelihood evaluation, the per-trajectory cost is
\begin{equation}\label{eq:C1}
  C_1  =  2 C_{\mathrm{LL}},
\end{equation}
where $C_{\mathrm{LL}}$ denotes the cost of a single log-likelihood
evaluation for the architecture in question.

Roughly speaking, the cost $C_{\mathrm{LL}}$ of a single log-likelihood evaluation
scales at most as $O(T^2)$ for Transformers~\cite{Vaswani2017}
and as $O(T)$ for recursive architectures such as RNNs~\cite{Elman1990,HochreiterSchmidhuber1997}
and state space models and Mamba~\cite{GuGoel2022,GuDao2023,SmithWarringtonLinderman2023};
in particular, no combinatorial overhead from enumerating non-Markovian histories enters either factor in \eqref{eq:total-cost}, the only $T$-dependence of $N$ being the Monte Carlo factor $\mathrm{Var} (\sigma )$.

 %----------------------------------------------------------------------
\subsection{Temporal coarse-graining}
\label{sec:implications-cg}
%----------------------------------------------------------------------

When our framework is applied to language models,
the backward process \eqref{eq:P-backward} evaluates the path probability
of the token sequence in reversed order.
For instance, given the forward sequence
$y_{1:4} = (\texttt{This}, \texttt{is}, \texttt{a}, \texttt{book})$,
the backward path probability is that of
$(\texttt{book}, \texttt{a}, \texttt{is}, \texttt{This})$,
which is extremely small under a model trained on natural language.
The entropy production is then dominated by this artifact of
token-level reversal, rather than by physically or semantically meaningful irreversibility that may reflect the structure of the real-world processes described by the text.

A natural approach to this issue is temporal coarse-graining:
reversing the order of \emph{blocks} of tokens rather than individual
tokens.
In this subsection, we restrict to the time-homogeneous case
\begin{equation}\label{eq:time-homogeneous}
  p_t = p, \qquad \Phi_t = \Phi \qquad \text{for all } t,
\end{equation}
with a common initial latent state $\tilde{h}_0 = h_0$.
This is the setting directly relevant for typical pre-trained LLMs,
in which a single set of model parameters is applied at every time
step.
Under \eqref{eq:time-homogeneous}, the forward path
probability~\eqref{eq:P-forward} becomes
\begin{equation}\label{eq:P-homogeneous}
  P(y_{1:T})
  = \prod_{t=0}^{T-1}
      p\bigl(y_{t+1} \mid \Phi(y_{1:t})\bigr),
\end{equation}
where $\Phi( \ ) \equiv h_0$.

To define coarse-grained reversal, we group consecutive tokens into
blocks
\begin{equation}
  y'_{t'} \equiv (y_{(t'-1)l+1}, \dots, y_{t'l})
\end{equation}
of length~$l$ (with $T = \tilde{T} l$)
and define the \emph{block-reversed token sequence}
\begin{equation}\label{eq:block-rev-def}
  \tilde{y}'_{1: \tilde{T} }
  \equiv
  \bigl(
    \underbrace{y_{( \tilde{T} -1)l+1}, \ldots, y_{ \tilde{T} l}}_{y'_{ \tilde{T} }},
    \underbrace{y_{ (\tilde{T} -2)l+1}, \ldots, y_{( \tilde{T} -1)l}}_{y'_{ \tilde{T} -1}},
    \ldots,
    \underbrace{y_{1}, \ldots, y_{l}}_{y'_{1}}
  \bigr),
\end{equation}
which concatenates blocks $y'_{ \tilde{T} }, y'_{ \tilde{T} -1}, \ldots, y'_1$
in reversed block order while preserving the token order within each
block.
For example, with $T = 6$, $l = 3$, and
$y_{1:6} = (a,b,c,d,e,f)$,
the token-level reversal is $(f, e, d, c, b, a)$,
while the block-level reversal is 
$\tilde{y}'_{1:2} =
    (d, e, f, a, b, c)$.

If one chooses the blocks at the level of sentences
or ``episodes'' (contiguous segments forming
semantically coherent units), the backward sequence reverses the order
of episodes but generates the tokens within each episode in the
forward order.
The intra-episode conditional probabilities would remain those of
natural language, and thus the entropy production would be governed by
inter-episode retrodiction, which may carry a more interpretable
signal.
This expectation is consistent with the GPT-2 demonstration in Section~\ref{sec:gpt2-demo}.

Since the model is time-homogeneous, the coarse-grained backward
path probability is simply the path probability of
the block-reversed sequence evaluated by the same model:
\begin{equation}\label{eq:cg-backward}
  P'_{\leftarrow}(y'_{ \tilde{T} :1})
  \equiv
  P(\tilde{y}'_{1: \tilde{T} }).
\end{equation}
The coarse-grained stochastic entropy production for a single
trajectory $y_{1:T} \sim P_{\to}$ is then
\begin{equation}\label{eq:cg-sigma}
  \sigma'(y_{1:T})
  \equiv
  \ln P(y_{1:T})
    - \ln P(\tilde{y}'_{1: \tilde{T} }),
\end{equation}
which is the difference in log-likelihood between the original text
and its block-reversed version, both evaluated by the same model
in a single forward pass each.

Since the block-reversal map is a bijection on $\mathcal{Y}^T$,
$P'_{\leftarrow}$ is a normalized distribution on $y_{1:T}$.
The coarse-grained entropy production
\begin{equation}\label{eq:cg-MC}
  \mathcal{S}'_y
  \equiv
  \mathbb{E}_{P_{\to}}[\sigma']
  =
  D_{\mathrm{KL}}\bigl(
    P_{\to}(y_{1:T}) \big\| P'_{\leftarrow}(y'_{ \tilde{T} :1})
  \bigr)
  \geq 0
\end{equation}
is non-negative as a KL divergence, and the integral fluctuation
theorem
$\mathbb{E}_{P_{\to}}[e^{-\sigma'}] = 1$
follows directly from the normalization of $P'_{\leftarrow}$.
The Monte Carlo estimator takes the same form
as~\eqref{eq:sigma-MC}:
$\mathcal{S}'_y \approx
  \frac{1}{N}\sum_{n=1}^{N} \sigma'(y_{1:T}^{(n)})$.

Concretely, $\sigma'(y_{1:T})$ is evaluated as follows.
Given a single sampled trajectory $y_{1:T}$,
the forward log-likelihood $\ln P(y_{1:T})$ is obtained by the
standard forward pass: feed $y_1, y_2, \ldots, y_T$ sequentially
into the model, compute the latent states
$h_t = \Phi(y_{1:t})$ at each step, and accumulate
$\sum_{t} \ln p(y_{t+1} \mid h_t)$.
For the second term, one constructs the block-reversed token sequence
$\tilde{y}'_{1: \tilde{T} }$~\eqref{eq:block-rev-def} and feeds it into the
\emph{same} model as if it were an ordinary input sequence:
the model computes new latent states
$\tilde{h}_1, \tilde{h}_2, \ldots$ by applying $\Phi$ to
successive prefixes of $\tilde{y}'_{1: \tilde{T} }$,
and one accumulates the log-likelihood of each token in
$\tilde{y}'_{1: \tilde{T} }$ conditioned on the corresponding
$\tilde{h}$, yielding $\ln P(\tilde{y}'_{1: \tilde{T} })$.
No sampling from the backward process is required;
the entire computation consists of two forward passes of the model
(one on the original sequence and one on its block-reversed version),
so the per-trajectory cost is
$C_1 = 2\,C_{\mathrm{LL}}$, the same as for the token-level
$\sigma$~\eqref{eq:C1}.
Setting $l=1$ reduces the block-reversed sequence to the fully
reversed sequence, and \eqref{eq:cg-backward} recovers the
token-level backward path probability~\eqref{eq:P-backward}.

The construction extends to variable-length blocks.
Let $S$ be a segmentation rule that partitions every
sequence $y_{1:T}$ into consecutive blocks
$B_1, \ldots, B_k$ of possibly unequal lengths summing to~$T$.
Given a forward sample $y_{1:T} \sim P$, one applies $S$
to obtain the blocks, concatenates them in reversed order
$R_S(y_{1:T}) \equiv B_k B_{k-1} \cdots B_1$,
and evaluates the log-likelihood of $R_S(y_{1:T})$ under the
same model; the stochastic entropy production is again
$\sigma'(y_{1:T})
  = \ln P(y_{1:T}) - \ln P\bigl(R_S(y_{1:T})\bigr)$,
 as in~\eqref{eq:cg-sigma}.
For $ P\bigl(R_S(y_{1:T})\bigr)$ to be a normalized distribution
(and hence for~\eqref{eq:cg-MC} to remain a valid KL divergence),
$R_S$ must be a bijection.
A natural strategy is to segment at every occurrence
of a distinguished delimiter token (e.g., a sentence-final period
or an end-of-sentence marker) and to require that $y_T$ itself be
such a delimiter.
In this case, the segmentation of the reversed sequence
is uniquely determined, and thus $R_S$ is 
a bijection on the restricted space conditioned on this terminal constraint.
We then replace $P$ with the conditional distribution $P_{\rm c}$, which leaves $\sigma'$ unchanged.
The fluctuation theorem for the coarse-grained entropy production is then satisfied:
\begin{equation}
\mathbb{E}_{P_{\rm c}} [e^{-\sigma' (y_{1:T})}] = 1.
\end{equation}

%======================================================================
\section{Proof-of-concept experiment with GPT-2}
\label{sec:gpt2-demo}
%======================================================================
 
As a demonstration of the estimation method formulated in Section~\ref{sec:sampling},
we evaluate the stochastic entropy production
for a pre-trained Transformer-based language model, GPT-2 (124M parameters)~\cite{Radford2019}.
Since GPT-2 uses time-independent parameters,
it satisfies the time-homogeneous
condition~\eqref{eq:time-homogeneous},
and the path probability takes
the form~\eqref{eq:P-homogeneous}.
For each text $y_{1:T}$, we compute two quantities:
the first is the token-level stochastic entropy production, 
here written as $\sigma_{\mathrm{token}}$,  and the second is the block-level stochastic entropy production
with variable-length blocks~\eqref{eq:cg-sigma}, 
written as $\sigma_{\mathrm{block}}$.
The segmentation and reversal are both performed at the level
of the token-ID sequence.
 
 In the following, we evaluate these quantities for the trajectory probabilities of GPT-2, using sampling from GPT-2 itself (Section~\ref{subsec:gpt2-generated}) and using fixed (externally prepared) text sets (Section~\ref{subsec:gpt2-fixed}).

\subsection{Monte Carlo sampling from GPT-2}
\label{subsec:gpt2-generated}

We generate sequences of $T=120$ tokens
from GPT-2 
(see Appendix~\ref{app:gpt2-boundary} for details).
Sampling is carried out by an explicit autoregressive loop
that draws one token at a time using the model's KV-cache,
rather than by the library's default \texttt{model.generate()}
method; this excludes any post-processing that deforms distributions, and ensures that exactly $T$~tokens are always
produced.
The forward log-likelihood can be accumulated from the same
logits used for sampling.

The path probability $P(y_{1:T})$
in~\eqref{eq:P-homogeneous} includes the $t=0$ factor
$p(y_1 \mid h_0)$, where $h_0 = \Phi( \ )$ is the
initial latent state.
We compute this factor by conditioning on GPT-2's
special \texttt{<|endoftext|>} token as the initial token.
Intuitively, this choice of the initial token means that any user-specified prompt is absent.
The same initial token is used for both the original
and reversed sequences, so that the common initial latent state
$\tilde{h}_0 = h_0$ is satisfied.

For the token-level entropy production
$\sigma_{\mathrm{token}}$,
we use each generated sequence of length~$T$.
For the block-level entropy production
$\sigma_{\mathrm{block}}$,
we truncate each sequence at the last sentence-final
punctuation token so that $y_{T'}$ is a delimiter,
as required by the bijection condition
(Section~\ref{sec:implications-cg}),
where $T' \le T$ denotes the truncated length.
Sequences containing no sentence-final punctuation are
excluded from the block-level analysis but retained
for the token-level one.
For each sample we compute the token-level
per-token entropy production
$\sigma_{\mathrm{token}}/T$
using the full sequence, and the block-level
$\sigma_{\mathrm{block}}/T'$
using the truncated sequence.
Additionally, for the sake of comparison,
we compute the token-level entropy production
$\sigma_{\mathrm{token}}(T') \equiv \sigma_{\mathrm{token}}(y_{1:T'})$
on the same truncated sequence used for the block-level analysis.

\begin{figure*}[htbp]
  \centering
  \includegraphics[width=0.8\linewidth]{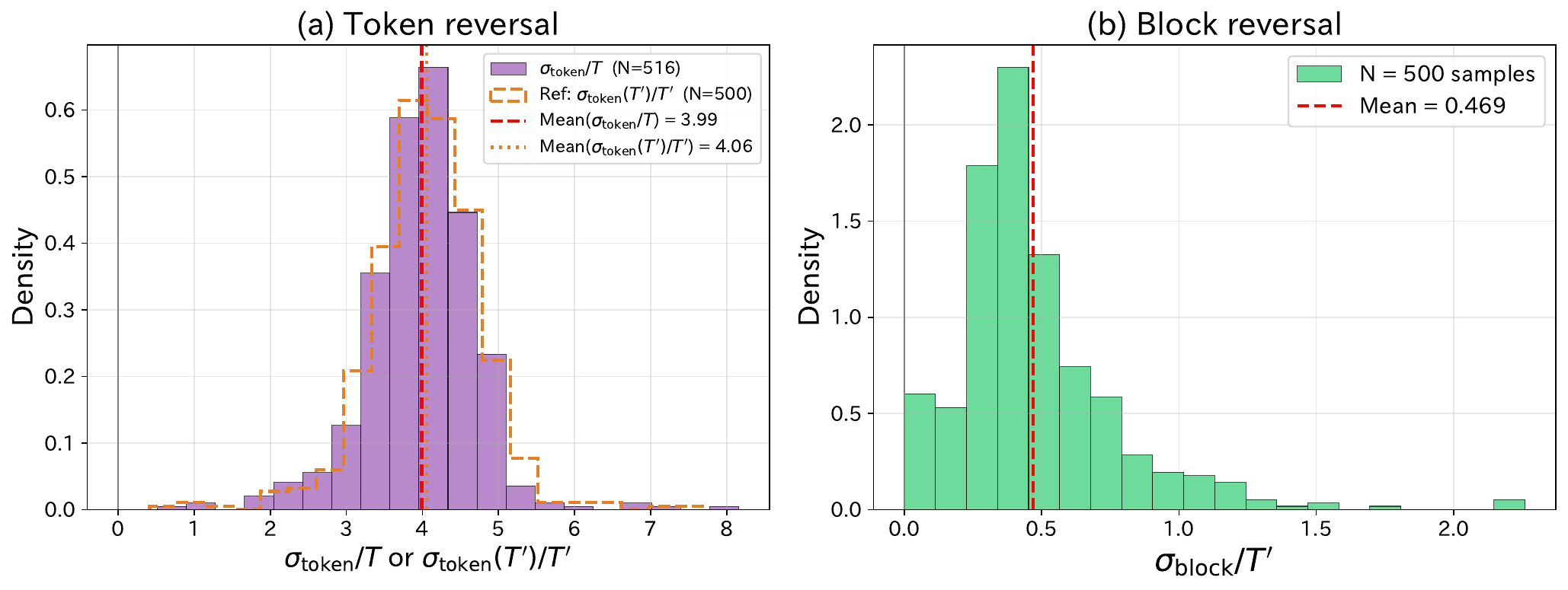}
 \caption{Distribution of the per-token stochastic entropy production
  for sequences of $T=120$ tokens sampled from GPT-2
  (no top-$k$ or nucleus truncation). 
  The temperature parameter is $\tau = 1$.
  (a)~Token-level reversal $\sigma_{\mathrm{token}}/T$,
  computed on full generated sequences (filled purple);
  the dashed orange step histogram shows the reference
  $\sigma_{\mathrm{token}}(T')/T'$,
  i.e.\ token-level reversal applied to the truncated
  sequences of length~$T'$.
  (b)~Block-level reversal $\sigma_{\mathrm{block}}/T'$,
  where each sequence is post-hoc truncated at the last
  sentence-final punctuation token (length~$T' \le T$).
  Dashed red lines indicate the sample means;
  the dotted orange line in~(a) indicates the mean of
  the reference distribution.
  We collect samples until $N = 500$ of them satisfy the
  bijection condition for block reversal (b). 
  Samples that fail this condition are excluded from the block-level analysis but retained for the token-level one, so the token-reversal count in~(a) is slightly larger (namely 516).
  Note the different horizontal scales between (a) and (b).}
  \label{fig:gpt2-histogram}
\end{figure*}

Figure~\ref{fig:gpt2-histogram} shows the resulting distributions of the stochastic entropy production.
The token-level values (a) are concentrated well above zero,
confirming the large irreversibility of token-order reversal:
the token-level entropy production is dominated by the artifact
of syntactic destruction --- a reversed token sequence such as
``\texttt{book a is This}'' receives extremely low probability
under GPT-2.
This dominance of the syntactic artifact in
$\sigma_{\mathrm{token}}$ motivates the block-level
coarse-graining, which has no counterpart in 
previous studies of forward--backward asymmetry in
LLMs~\cite{PapadopoulosWengerHongler2024,YuEtAl2025}.
In fact, the block-level values shown in Figure~\ref{fig:gpt2-histogram}  (b) are much smaller,
which is consistent with the discussion in
Section~\ref{sec:implications-cg}.
Note that the reference distribution
$\sigma_{\mathrm{token}}(T')/T'$,
overlaid in (a), has  almost the same mean as
that of $\sigma_{\mathrm{token}}/T$,
indicating that the much smaller block-level values in
(b) are mainly due to the coarser reversal 
rather than the truncation from $T$ to~$T'$.

We also numerically observe that, for both token reversal and block
reversal, the Monte Carlo estimate of the entropy production
 converges to a positive value by around $N = 500$;
see Appendix~\ref{app:gpt2-ft} for supplemental numerical
results, including a verification of the integral fluctuation
theorem for token reversal, as a consistency check.

The token-level Monte Carlo sampling without truncation (the filled purple in Figure~\ref{fig:gpt2-histogram} (a)) is taken over
the true distribution $P(y_{1:T})$ of GPT-2, which exactly fits our general theoretical description.
For the block-level reversal, however,
the estimated quantity deviates from the true distribution
in two respects.
First, sequences lacking any sentence-final punctuation
are excluded; this exclusion rate is about 3\%
in our experiment.
Second, each retained sequence is truncated at the position of
its last delimiter.
Denoting the truncation point $T' (y_{1:T})$,
the block-level Monte Carlo estimator converges to
$\mathbb{E}_{y_{1:T} \sim P'} \left[ \sigma_{\mathrm{block}}(y_{1:T'}) / T' (y_{1:T}) \right]$, where $P'$ is the conditional distribution under the above-mentioned exclusion.
Therefore, rigorously speaking,
our Monte Carlo estimate for the block-reversal case
deviates from the ideal theoretical setting described
in Section~\ref{sec:implications-cg}.
However, we expect this deviation to be small, given that the analogous deviation for token-level reversal is minor (compare the filled purple and dashed orange distributions in Figure~\ref{fig:gpt2-histogram}(a)).

In addition, the dependence of the coarse-grained entropy production on the block scale, for both fixed-length blocks and $k$-sentence superblocks, is examined in Appendix~\ref{app:block-scale-dependence} (Fig.~\ref{fig:gpt2-block-scale}).

\subsection{Evaluation of externally prepared texts }
\label{subsec:gpt2-fixed}

An important challenge is to clarify the meaning and
implications of the entropy production at both the
token-reversal and block-reversal levels, when applied
to real language models.
As a first step in this direction, we present an experiment
in which we evaluate the stochastic entropy production of
GPT-2 on externally prepared texts, rather than on sequences
generated by GPT-2 itself.
This is still meaningful as a probe of GPT-2, because the
stochastic entropy production quantifies the irreversibility
of a \emph{single} realized trajectory.
A practical difficulty, however, is how to prepare test
texts systematically while avoiding arbitrary choices by the
experimenter.

To address this issue, we use multiple input-text generators and
examine whether the observed causal/non-causal difference depends on
the model used to prepare the texts. Specifically, we submit the same
fixed prompt to five language models: Claude Opus 4.6, Claude Fable 5, Gemini 3.1 Pro, GPT-5.4 Pro, and GPT-5.6 Pro. For each
generator, we prepare the following two sets of English-language
texts, each containing 100 four-sentence samples:

\begin{itemize}
  \item \emph{Causal texts.}
    Short narratives in which the sentences describe
    a temporally ordered chain of events
    (e.g., ``The glass slipped from Alice's hand. The glass fell to the floor. The glass broke into many pieces. The glass left a wet stain on the tile floor.'').
    Reversing the sentence order yields a description
    in which effects precede their causes.
  \item \emph{Non-causal texts.}
    Collections of independent factual statements whose ordering
    carries no temporal or causal implication
    (e.g., ``The library lent a heavy atlas. The library hosted a chess tournament. The library displayed old framed photographs. The library received a donated wooden shelf from a volunteer.'').
    Reversing the sentence order leaves the meaning essentially
    unchanged.
\end{itemize}
 
The two examples above are included as fixed entries in every
corpus. 
To make the surface form of the two sets more comparable, the prompt imposes a common format: all four sentences in each text begin with the same noun phrase, every sentence is written in the simple past tense, and pronouns, demonstratives, and inter-sentence connectives are excluded.
The complete text sets, prompts, analysis code, and raw
numerical results are provided in the GitHub repository (see the
Data Availability statement).

 For every text, both the token-reversal
quantity $\sigma_{\mathrm{token}}/T$ and the sentence-block-reversal
quantity $\sigma_{\mathrm{block}}/T$ are evaluated using the same
GPT-2 model; the five models listed above are used only to generate the
input texts.

\begin{figure*}[t]
  \centering
  \includegraphics[width=\textwidth]{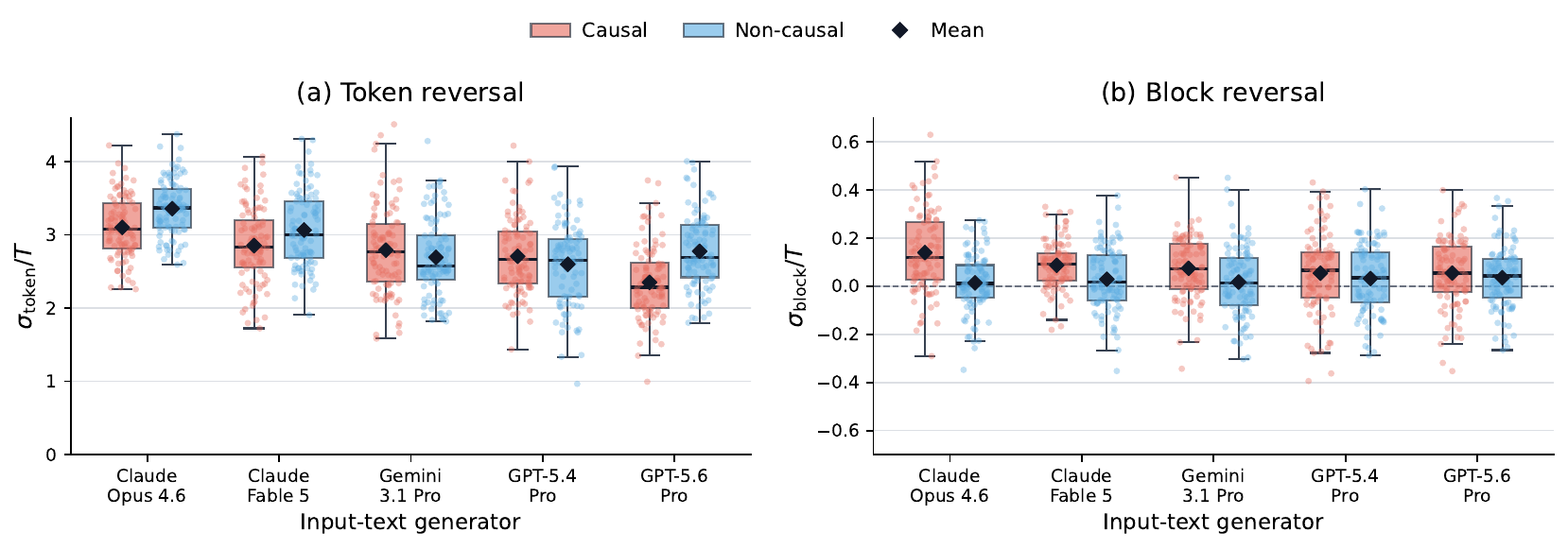}
  \caption{
  Distributions of the stochastic entropy production  evaluated using GPT-2 for the fixed
  texts generated by the five input-text models.
  (a) Token-level reversal $\sigma_{\mathrm{token}}/T$.
  (b) Sentence-block reversal $\sigma_{\mathrm{block}}/T$.
  For each generator, red and blue denote the 100 causal and 100
  non-causal texts, respectively.
  All individual observations are shown with horizontal jitter, and no observations are removed as outliers.
  Boxes span the interquartile range, horizontal lines indicate the
  median, whiskers extend to the most extreme observation within
  1.5 times the interquartile range from the box, and black diamonds
  indicate the mean.
  Since the input texts  are not sampled from GPT-2's distribution, the averages of $\sigma$ are not guaranteed to be non-negative.
  Note the different vertical scales in the two panels.
  }
  \label{fig:fixed-text-distributions}
\end{figure*}

The resulting distributions are shown in
Fig.~\ref{fig:fixed-text-distributions}. 
The token-level difference between the two categories varies strongly
across input-text generators, whereas the block-level difference is consistently larger for the causal texts.

For each generator and reversal scheme, we compare the causal and
non-causal samples using a two-sided asymptotic Mann--Whitney $U$ test under an independence approximation. We quantify the direction and magnitude
of the difference by the rank-biserial correlation
\begin{equation}
 r \equiv \frac{2U_{\mathrm{C}}}{n_{\mathrm{C}}n_{\mathrm{NC}}}-1
   = \frac{N_{>}-N_{<}}{n_{\mathrm{C}}n_{\mathrm{NC}}},
   \label{def_biserial_r}
\end{equation}
where $U_{\mathrm{C}}$ is the Mann--Whitney statistic for the causal
sample, $n_{\mathrm{C}}$ and $n_{\mathrm{NC}}$ are the causal and
non-causal sample sizes, and $N_{>}$ ($N_{<}$) is the number of
cross-group pairs for which the causal value is larger (smaller) than
the non-causal value. Tied pairs contribute one half to
$U_{\mathrm{C}}$ and zero to $N_{>}-N_{<}$. Thus, $-1\leq r\leq 1$:
$r>0$ means that the causal value tends to be larger, $r<0$ means that
the non-causal value tends to be larger, and $r=0$ indicates no net
pairwise ordering preference.

\begin{figure*}[t]
  \centering
  \includegraphics[width=\textwidth]{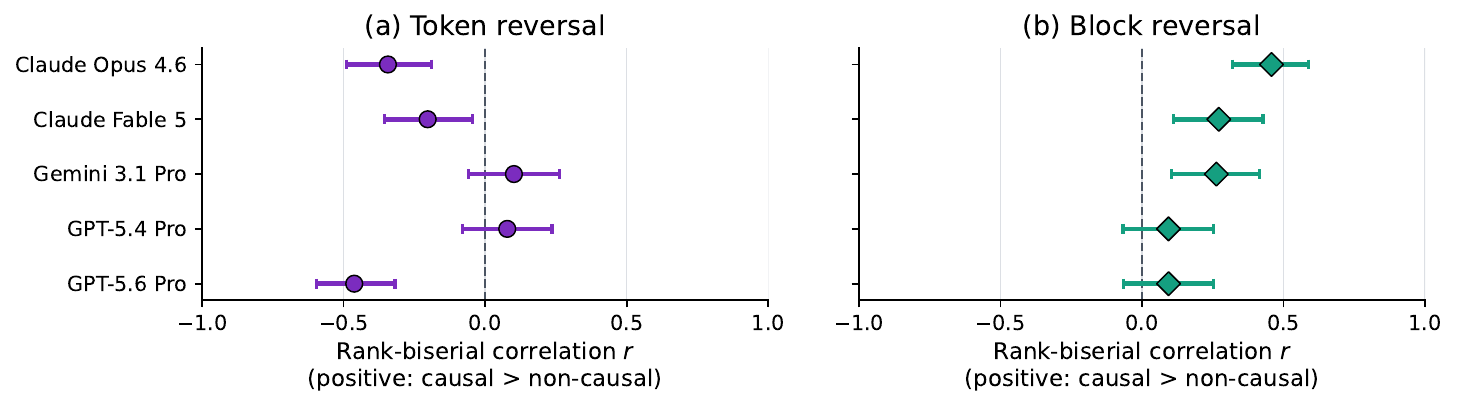}
  \caption{
  Cross-generator comparison of the causal/non-causal difference in
  the per-token stochastic entropy production evaluated by GPT-2.
  The labels on the vertical axis indicate the language model used to
  generate each input-text corpus.
  (a) Token-level reversal $\sigma_{\mathrm{token}}/T$.
  (b) Sentence-block reversal $\sigma_{\mathrm{block}}/T$.
  Each point is the rank-biserial correlation $r$ defined in Eq.~\eqref{def_biserial_r} for
  100 causal and 100 non-causal texts. Positive values indicate that the causal texts
  tend to have larger values than the non-causal texts. Error bars show
  pointwise 95\% percentile bootstrap confidence intervals obtained by
  resampling independently within the two text categories ($B=10^4$).
  All observations are included; no outlier removal is performed.
  }
  \label{fig:fixed-text-cross-model}
\end{figure*}

Figure~\ref{fig:fixed-text-cross-model} reveals a qualitative
difference between the two reversal schemes. For token reversal, the
causal/non-causal difference depends strongly on the input-text
generator: three of the five models show significantly smaller
$\sigma_{\mathrm{token}}/T$ for the causal texts ($r=-0.46$ to
$-0.20$), while the other two show small positive estimates whose
$95\%$ confidence intervals include zero.
 This variation, which far exceeds the sampling uncertainty, is consistent with sensitivity to
generator-specific stylistic or structural properties, although the
present experiment does not identify which such properties drive it.

For block reversal, by contrast, every significant difference (three of the five generators) is in the direction of larger
$\sigma_{\mathrm{block}}/T$ for the causal texts ($r=+0.26$ to
$+0.46$), which is the direction expected by construction. The
remaining two estimates are positive but small and not significant,
and no generator shows a significant difference in the opposite
direction.
That is, the magnitude of the block-level effect varies across
generators, but its sign always agrees
with the expected direction. Detailed values of $U$, $p$, and $r$ are
reported in Appendix~\ref{app:fixed-texts}.

These results favor block reversal over token reversal as a probe of
inter-sentence ordering in the present setting: its response to the
causal/non-causal manipulation is consistent in sign and interpretable
in terms of event ordering.

In Appendix~\ref{app:qwen3-fixed-text} (Figs.~\ref{fig:qwen3-fixed-text-distributions} and \ref{fig:qwen3-fixed-text-effects}), we repeat the experiment with the same five corpora but with a
substantially larger evaluator, Qwen3-4B-Base (4.0 billion parameters). The block-level
difference is then statistically significant for all five generators
(largest $p = 0.045$, Table~\ref{tab:fixed-text-statistics2}; all five
individual unadjusted tests are significant). By contrast, the
token-level difference remains generator-dependent, with significant
effects in both directions. This re-evaluation thus supports the
GPT-2 results with improved statistical resolution.

 We do not claim, however, that
these results establish $\sigma_{\mathrm{block}}/T$ as a general
quantitative measure of causal structure. The causal/non-causal labels
are defined by the generation prompt rather than by a formal criterion,
and correlations between these labels and generator-specific stylistic
features cannot be completely excluded. A more rigorous assessment
would require a prompt-independent definition of causal ordering;
constructing such a definition remains an important direction for
future work.

Finally, our ``causal'' and ``non-causal''
categories do not rigorously distinguish mere temporal ordering
from genuine causal dependence --- a distinction that
is inherently difficult even in
principle~\cite{JamesBarnettCrutchfield2016}.
Recent work  has shown that
LLMs in fact tend to confound these two relations~\cite{MilianiEtAl2025}.

%======================================================================
\section{Linear Gaussian case: Kalman innovation representation}
\label{sec:linear-gaussian-example}
%======================================================================
 
As an analytically tractable demonstration of the general framework,
we specialize to the linear Gaussian case, where the autoregressive
model coincides with the innovation representation of the steady-state
Kalman filter~\cite{Kalman1960,AndersonMoore1979,LindquistPicci2015}.
We obtain an analytical expression for the entropy production
$\mathcal{S}_y$ by introducing the innovation reversal
matrix~$\mathcal{R}$,
which is also verified by Monte Carlo sampling.

We note that Refs.~\cite{MitterNewton2005,Newton2006,Newton2007,Newton2008}
studied entropy production associated with the information flow
between the signal process~$x_t$ and the observation process~$y_t$
in continuous-time Kalman--Bucy and nonlinear filters.
In contrast, the quantity computed below is the KL divergence
between the forward and time-reversed path measures of the
observation sequence~$y_t$ alone, with no reference to the
underlying state~$x_t$.

Separately, thermodynamic aspects of linear Gaussian systems
involving Kalman filtering have also been explored from
complementary perspectives in~\cite{HorowitzSandberg2014,SandbergDelvenneNewtonMitter2014,MatsumotoSagawa2018,KumasakiTojoSagawaFuno2026}.
We note that the entropy production in continuous-time linear Gaussian systems
has also been analyzed in the context of multivariate
Ornstein--Uhlenbeck
processes~\cite{GodrecheLuck2019,LandiTomeOliveira2013,GilsonTagliazucchiCofre2023}.

%----------------------------------------------------------------------
\subsection{Setup}
%----------------------------------------------------------------------
 
Consider a linear Gaussian process with state dimension~$n_x$
and observation dimension~$n_y$:
\begin{align}
  x_{t+1} &= A  x_t + w_t, \qquad w_t \sim \mathcal{N}(0, Q), \label{eq:state}\\
  y_t     &= C  x_t + v_t, \qquad v_t \sim \mathcal{N}(0, R), \label{eq:obs}
\end{align}
where we assume, for simplicity, $A \in \mathbb{R}^{n_x \times n_x}$, $C \in \mathbb{R}^{n_y \times n_x}$,
$Q \in \mathbb{R}^{n_x \times n_x}$ (positive semi-definite),
and $R \in \mathbb{R}^{n_y \times n_y}$ (positive definite) are all time-independent.
The noise sequences $\{w_t\}$ and $\{v_t\}$ are mutually independent
and independent across time.
 
We assume that the Kalman filter~\cite{Kalman1960} has reached steady state
(see, e.g., \cite{AndersonMoore1979} for standard results used below).
Then, the following relevant quantities are all time-independent:  the prediction error covariance $P$ is given by the positive semi-definite solution of the Riccati equation 
\begin{equation}
P = A\bigl(P - P C^\top (C P C^\top + R)^{-1} C P\bigr) A^\top + Q,
\end{equation}
 the innovation covariance is given by $S = C P C^\top + R$, and the Kalman gain is given by $K = P C^\top S^{-1}$.
 This stationary assumption concerns the
filter parameters only; by itself it does not imply that the finite-time
observation path probability is stationary.

 To connect with the general framework of Section~\ref{sec:setup},
note that the true state $x_t$ in~\eqref{eq:state}--\eqref{eq:obs} has no counterpart in our
framework: it appears neither in the latent state $h_t$ nor in
the emission kernel $p_t$.
Instead, the latent state is identified as
$h_t = \hat{x}_{t+1|t}$, the one-step-ahead predicted estimate
(with $h_0 = \hat{x}_{1|0}$), which is a deterministic function of
past observations $y_1, \dots, y_t$, and the emission kernel is
$p_t(y_{t+1} \mid h_t) = \mathcal{N}(C h_t,  S)$.
The generative process of Section~\ref{sec:gen} thus produces
$y_{1:T}$ entirely through the deterministic--stochastic loop between
$h_t$ and $y_{t+1}$, with $x_t$ absent; this is the sense in which
we regard the Kalman filter as a generative model
(see also Section~\ref{sec:examples} and Table~\ref{tab:examples}).
 
%----------------------------------------------------------------------
\subsection{Forward process}
\label{sec:gen}
%----------------------------------------------------------------------
 
The generative process produces a sequence $y_1, y_2, \dots, y_T$ as follows.
Fix an initial predicted state estimate $\hat{x}_{1|0}$ (e.g.\ $\hat{x}_{1|0}=0$).
Then, for $t = 1, 2, \dots, T$:
 
\begin{enumerate}
  \item \textit{Draw an innovation.}
    Sample $e_t \sim \mathcal{N}(0, S)$ independently.
  \item \textit{Generate the observation.}
    Set
    \begin{equation}
      y_t = C \hat{x}_{t|t-1} + e_t.
      \label{eq:gen_y}
    \end{equation}
  \item \textit{Kalman filter update.}
    Compute the next predicted state estimate deterministically:
    \begin{align}
      \hat{x}_{t|t}   &= \hat{x}_{t|t-1} + K  ( y_t - C \hat{x}_{t|t-1}), \label{eq:update}\\
      \hat{x}_{t+1|t} &= A \hat{x}_{t|t}. \label{eq:predict}
    \end{align}
\end{enumerate}
 
\noindent
Since $e_t = y_t - C \hat{x}_{t|t-1}$ and $\hat{x}_{t|t-1}$ is a deterministic
function of $y_1, \dots, y_{t-1}$, the conditional distribution is
\begin{equation}
  y_t \mid y_{1:t-1}  \sim  \mathcal{N} \bigl(C \hat{x}_{t|t-1},  S\bigr).
  \label{eq:cond}
\end{equation}
By the innovation decomposition~\cite{AndersonMoore1979}, the joint density of the forward
path is
\begin{equation}
\begin{split}
  &P_{\to}(y_{1:T})
  = \prod_{t=1}^{T} \mathcal{N} \bigl(e_t;  0,  S\bigr) \\
  &{} \ = \frac{1}{(2\pi)^{Tn_y/2} (\det S)^{T/2}}
    \exp \Biggl(-\frac{1}{2}\sum_{t=1}^{T} e_t^\top S^{-1} e_t\Biggr).
\end{split}
  \label{eq:pF}
\end{equation}
 
This coincides with the marginal distribution of $y_{1:T}$ obtained by integrating
out the states $x_{1:T}$ in the original dynamics, for a
compatible initial distribution of $x_1$.  Hence, for any fixed initial predictor
$\hat x_{1|0}$, the forward path vector $y_{1:T}=(y_1,\dots,y_T)$ is
Gaussian:
\begin{equation}
  P_{\to} (y_{1:T}) =  \mathcal{N}(y_{1:T};m,\Sigma),
  \label{eq:joint_gauss}
\end{equation}
where $m$ and $\Sigma$ are the mean vector and the covariance matrix, determined by the update rule described above.
If $\hat x_{1|0}=0$, then $m=0$; this is always assumed in the following.
 
Combining the filter update~\eqref{eq:update} and
prediction~\eqref{eq:predict} and iterating from
$\hat{x}_{1|0}=0$, one obtains
the causal moving-average (innovation) representation~\cite{AndersonMoore1979,LindquistPicci2015}
\begin{equation}\label{eq:MA-rep}
  y_t
  = \sum_{k=1}^{t} H_{t-k} e_k,
\end{equation}
where the impulse-response coefficients are
\begin{equation}\label{eq:H-coeff}
  H_0 \equiv I_{n_y}, \qquad
  H_l \equiv C A^{l} K \quad (l \geq 1).
\end{equation}
With the stacked vectors
$\mathbf{y} \equiv (y_1^\top, \dots, y_T^\top)^\top$ and
$\mathbf{e} \equiv (e_1^\top, \dots, e_T^\top)^\top$,
\eqref{eq:MA-rep} reads
\begin{equation}
\mathbf{y} = \mathcal{H}\mathbf{e},
\end{equation}
where $\mathcal{H} \in \mathbb{R}^{Tn_y \times Tn_y}$ is the block
lower-triangular matrix with $(i,j)$ block $H_{i-j}$ for $i \geq j$
and zero otherwise.
The covariance matrix of $\mathbf{e}$  is given by 
\begin{equation}
\mathbb{E}_{P_{\to}} \Bigl[ \mathbf{e} \ \mathbf{e}^\top \Bigr] = I_T \otimes S,
\end{equation}
where $I_T$ is the identity matrix of the auxiliary $T$-dimensional system that encodes the labels of time, and $\otimes$ denotes the tensor product between the auxiliary space and the original $n_y$-dimensional space. 
The path
covariance matrix is then given by
\begin{equation}\label{eq:Sigma-analytic}
  \Sigma
  =
  \mathbb{E}_{P_{\to}} \Bigl[ \mathbf{y} \ \mathbf{y}^\top \Bigr]
  = \mathcal{H} (I_T \otimes S) \mathcal{H}^\top.
\end{equation}

%----------------------------------------------------------------------
\subsection{Backward process}
\label{sec:rev}
%----------------------------------------------------------------------

Following the general protocol described in Section~\ref{sec:backward},  
 we run the generative mechanism of Section~\ref{sec:gen}
on the backward sequence $\tilde y_{1:T}$:
\begin{enumerate}
  \item \textit{Draw an innovation.}
    Sample $\tilde e_s^B \sim \mathcal{N}(0, S)$ independently.
  \item \textit{Generate the observation.}
    Set
   \begin{equation}
   \tilde y_s = C \hat{x}^B_{s|s-1} + \tilde  e_s^B.
  \label{eq:eR}
\end{equation}
  \item \textit{Kalman filter update.}
    Compute the next predicted state estimate deterministically:
    \begin{align}
  \hat{x}^B_{s|s}   &= \hat{x}^B_{s|s-1} + K  ( \tilde y_s - C \hat{x}^B_{s|s-1}),
  \label{eq:updateR}\\
  \hat{x}^B_{s+1|s} &= A \hat{x}^B_{s|s}.
  \label{eq:predictR}
\end{align}
\end{enumerate}
We set the initial condition as $\hat{x}^B_{1|0} = \hat{x}_{1|0}$.

Since each innovation in the generative process is an independent draw from
$\mathcal{N}(0,S)$, the backward path density is the product of the densities
$\mathcal{N}(\tilde e_s^B; 0, S)$ evaluated at the values~\eqref{eq:eR}:
\begin{equation}
\begin{split}
  &P_{\leftarrow}(\tilde y_{1:T})
   =  \prod_{s=1}^{T} \mathcal{N} \bigl(\tilde e_s^B;  0,  S\bigr) \\
   &{} \ =  \frac{1}{(2\pi)^{Tn_y/2} (\det S)^{T/2}}
    \exp \Biggl(-\frac{1}{2}\sum_{s=1}^{T} (\tilde e_s^B)^\top S^{-1}  \tilde e_s^B\Biggr).
    \end{split}
  \label{eq:pR}
\end{equation}
Here, we do not assume \eqref{eq:assume_final} but assume that $\tilde y_1$ is sampled from an independent Gaussian distribution $\mathcal{N}(C \hat x_{1|0}, S)$.
Since the forward and backward path probabilities, \eqref{eq:pF} and \eqref{eq:pR}, share the same functional form,
\begin{equation}
  P_{\leftarrow}(\tilde{\mathbf{y}})  = \mathcal{N}(\tilde{\mathbf{y}}; 0,\Sigma).
    \label{eq:pR0}
  \end{equation}

 We now consider the particular  event that the backward-trajectory realization is the exact time-reversal of the forward trajectory,
 \begin{equation}
  \tilde y_s  =  y_{T-s+1}, \qquad s = 1, 2, \dots, T.
  \label{eq:z_def}
\end{equation}
With the time-reversal (permutation) matrix
\begin{equation}
  J =
  \begin{pmatrix}
    0      & \cdots & 0      & I_{n_y} \\
    0      & \cdots & I_{n_y} & 0      \\
    \vdots & \iddots & \vdots & \vdots \\
    I_{n_y} & \cdots & 0      & 0
  \end{pmatrix}
   \in  \mathbb{R}^{Tn_y \times Tn_y},
  \label{eq:J_def}
\end{equation}
we have $J\mathbf{y} = (y_T^\top, \dots, y_1^\top)^\top$.
Note that $J = J^\top = J^{-1}$ and $\det J = \pm 1$.
By substituting $\tilde{\mathbf{y}} = J \mathbf{y}$ into \eqref{eq:pR0},
\begin{equation}
  P_{\leftarrow}(J \mathbf{y})
  = \mathcal{N}(\mathbf{y}; 0,\widetilde{\Sigma}),
  \label{eq:pR_gauss}
\end{equation}
with covariance
\begin{equation}
  \widetilde{\Sigma} = J \Sigma  J.
  \label{eq:Sigma_tilde}
\end{equation}
Therefore, the sampling of observed trajectories in the backward process can be mapped to the sampling in the forward process, just by applying $J$ as above.

We now introduce $(e^B_1, e^B_2, \dots, e^B_T)$ via $J \mathbf{y} = \mathcal{H} \mathbf{e}^B$, that is,
\begin{equation}\label{def_e_B_new}
\mathbf{e}^B \equiv \mathcal{H}^{-1} J \mathbf{y}.
\end{equation}
Note that $\mathcal{H}$ is always invertible because the diagonal blocks are given by $H_0 = I_{n_y}$.
From $ \mathbf{y} = \mathcal{H} \mathbf{e}$, $\mathbf{e}^B$ is related to the original forward innovation as
\begin{equation}
\mathbf{e}^B  = \mathcal{R} \mathbf{e},
\end{equation}
where we introduced the \emph{innovation reversal matrix}
\begin{equation}\label{eq:new-R-def}
   \mathcal{R}
   \equiv 
  \mathcal{H}^{-1} J \mathcal{H}.
\end{equation}

The backward path density~\eqref{eq:pR} defines a probability measure
$P_{\leftarrow}$ on the space of observation sequences via the
generative process~\eqref{eq:eR}--\eqref{eq:predictR}, in which the
innovations $\tilde{e}_s^B$ are independently sampled from
$\mathcal{N}(0, S)$ and causally produce the observations
$\tilde{y}_s$.  To compute the entropy production
$\sigma = \ln P_{\to}(\mathbf{y}) - \ln P_{\leftarrow}(J\mathbf{y})$
for a given forward realization $\mathbf{y}$, one needs the value of
$P_{\leftarrow}(J\mathbf{y})$.  This value is obtained without
sampling from the backward process: one substitutes
$\tilde{y}_s = y_{T-s+1}$ into the density~\eqref{eq:pR}, which
requires the innovations $\tilde{e}_s^B$ as a function of the
observations.  To this end, one reads the
relation~\eqref{eq:eR} in the reverse direction: given an observation
$\tilde{y}_s$, one extracts the innovation as
$\tilde{e}_s^B = \tilde{y}_s - C \hat{x}^B_{s|s-1}$, and then
updates the state via~\eqref{eq:updateR}--\eqref{eq:predictR}.
Substituting the particular values $\tilde{y}_s = y_{T-s+1}$
and writing $e_s^B$ for the resulting innovations, one obtains,
starting from $\hat{x}^B_{1|0} = \hat{x}_{1|0}$, for
$s = 1, 2, \dots, T$,
\begin{align}
  e_s^B          &= y_{T-s+1} - C \hat{x}^B_{s|s-1},
  \label{eq:eB_extract} \\
  \hat{x}^B_{s|s}   &= \hat{x}^B_{s|s-1} + K e_s^B,
  \label{eq:updateB} \\
  \hat{x}^B_{s+1|s} &= A \hat{x}^B_{s|s}.
  \label{eq:predictB}
\end{align}
This is precisely the operation encoded by $\mathcal{H}^{-1}$ in
$\mathbf{e}^B = \mathcal{H}^{-1} J \mathbf{y}$: it runs the Kalman
filter on the time-reversed observation sequence
$(y_T, y_{T-1}, \dots, y_1)$ and deterministically extracts the
innovation at each step.  
We note that $\hat x^B_{s|s-1}$ is not  the time-reversal of $\hat x_{t|t-1}$ even when $\tilde{y}_s = y_{T-s+1}$, as mentioned for the general case in Section~\ref{sec:backward_ep}.

For the particular realization $\tilde{y}_s = y_{T-s+1}$, the
quantities $\tilde{e}_s^B$ and $e_s^B$ take the same numerical value,
since they are computed by the same recursion applied to the same
input sequence; the distinction lies solely in the probability distributions.  In the
backward generative process, $\tilde{e}_s^B$ is the independent random
input that produces $\tilde{y}_s$ via~\eqref{eq:eR};
under $P_{\leftarrow}$, it is i.i.d.\ $\mathcal{N}(0, S)$ by
construction.  By contrast, $e_s^B$ is deterministically
extracted from the forward trajectory $y_{1:T}$
via~\eqref{eq:eB_extract}--\eqref{eq:predictB};
under $P_{\to}$, $e_s^B = [\mathcal{R} \mathbf{e}]_s$ is in general
correlated across time steps and not identically distributed.  It is
precisely this mismatch between the i.i.d.\ statistics assumed by the
backward model and the actual statistics of $e_s^B$ under $P_{\to}$
that gives rise to a nonzero entropy production.

\subsection{Analytical expression for entropy production}

In general, the KL divergence between two Gaussians
$\mathcal{N}(\mu_1,\Sigma_1)$ and $\mathcal{N}(\mu_2,\Sigma_2)$ is
(see, e.g., \cite{CoverThomas2006})
\begin{equation}
\begin{split}
  &D_{\mathrm{KL}} \bigl(\mathcal{N}(\mu_1,\Sigma_1) \big\| \mathcal{N}(\mu_2,\Sigma_2)\bigr) \\
  &{} \ = \frac{1}{2}\Bigl[
    \operatorname{tr} \bigl(\Sigma_2^{-1} \Sigma_1\bigr)
    + (\mu_1 - \mu_2)^\top \Sigma_2^{-1} (\mu_1 - \mu_2) \\
    &{} \quad \quad \quad
    - d
    + \ln\frac{\det \Sigma_2}{\det \Sigma_1}
  \Bigr],
  \end{split}
  \label{eq:kl_gauss_general}
\end{equation}
where $d$ is the dimension of the random vector.

We now evaluate each term with $\mu_1 = \mu_2 = 0$, $\Sigma_1 = \Sigma$,
$\Sigma_2 = \widetilde{\Sigma} = J\Sigma J$ \eqref{eq:Sigma_tilde}, and $d = Tn_y$.
Since both distributions have zero mean, the quadratic
mean-difference term vanishes.
Since $J$ is an orthogonal matrix ($J^\top J = I$),
\begin{equation}
  \det \widetilde{\Sigma}
  = \det(J\Sigma J)
  = (\det J)^2 \det \Sigma
  = \det \Sigma.
\end{equation}
Therefore
\begin{equation}
  \ln \frac{\det \widetilde{\Sigma}}{\det \Sigma} = 0.
  \label{eq:det_cancel}
\end{equation}
The inverse of $\widetilde{\Sigma}$ is $\widetilde{\Sigma}^{-1}   = (J\Sigma J)^{-1}   = J^{-1} \Sigma^{-1} J^{-1}   = J \Sigma^{-1} J$,  where we used $J^{-1} = J$. Thus,
\begin{equation}
  \operatorname{tr} \bigl(\widetilde{\Sigma}^{-1} \Sigma\bigr)
  = \operatorname{tr} \bigl(J \Sigma^{-1} J \Sigma\bigr).
  \label{eq:trace_term}
\end{equation}
 Combining~\eqref{eq:det_cancel} and~\eqref{eq:trace_term} into~\eqref{eq:kl_gauss_general},  we obtain
\begin{equation}
  D_{\mathrm{KL}}(P_{\to} (\mathbf{y} ) \|  P_{\leftarrow} (J\mathbf{y})) 
  \quad
  = \frac{1}{2}\Bigl[
    \operatorname{tr} \bigl(J \Sigma^{-1} J \Sigma\bigr) - T n_y
  \Bigr].
  \label{eq:kl_final}
\end{equation}

We next rewrite~\eqref{eq:kl_final} in terms of $\mathcal{R}$~\eqref{eq:new-R-def}.
From the factorization~\eqref{eq:Sigma-analytic} and by letting $\mathcal{G} \equiv \mathcal{H}^{-1}$, 
$\Sigma^{-1}
  = \mathcal{G}^{ \top} (I_T \otimes S^{-1}) \mathcal{G}$.
Substituting into the trace term in~\eqref{eq:kl_final}
and using the cyclic property of the trace together with
$J = J^\top = J^{-1}$,
\begin{align}
  \operatorname{tr} \bigl(J \Sigma^{-1} J \Sigma\bigr)
  &= \operatorname{tr} \bigl(
       (I_T \otimes S^{-1}) 
       \underbrace{\mathcal{G} J \mathcal{H}}_{\mathcal{R}} 
       (I_T \otimes S) 
       \underbrace{\mathcal{H}^\top J \mathcal{G}^{ \top}}_{\mathcal{R}^\top}
     \bigr).
  \label{eq:new-trace-via-R}
\end{align}
Therefore, we obtain
\begin{equation}\label{eq:new-kl-R}
\begin{split}
  &D_{\mathrm{KL}}(P_{\to} (\mathbf{y} ) \|  P_{\leftarrow} (J\mathbf{y})) \\
  &\quad = \frac{1}{2}\Bigl[
    \operatorname{tr} \bigl(
      (I_T \otimes S^{-1}) \mathcal{R} 
      (I_T \otimes S) \mathcal{R}^\top
    \bigr)
    - T n_y
  \Bigr].
  \end{split}
\end{equation}

Meanwhile, since $\mathbf{e}^B = \mathcal{R} \mathbf{e}$
and $\mathbf{e} \sim \mathcal{N}(0, I_T \otimes S)$
under the forward measure $P_{\to}$,
each $e_s^B = \sum_{k=1}^{T}[\mathcal{R}]_{sk} e_k$
has covariance
\begin{equation}
\begin{split}
\Sigma_s^B
 &\equiv
  \mathbb{E}_{P_\to} \Bigl[ e_s^B (e_s^B)^\top  \Bigr]
  =  \sum_{k,k'=1}^{T} [\mathcal{R}]_{sk}  \underbrace{{\mathbb{E}_{P_{\to}} [e_k e_{k'}^\top]}}_{S\delta_{kk'}} [\mathcal{R}]_{sk'}^\top \\
  &=  \sum_{k=1}^{T} [\mathcal{R}]_{sk}  S  [\mathcal{R}]_{sk}^\top,
  \end{split}
 \end{equation}
 where the subscripts run over the auxiliary subspace of the time indexes.
 Thus,  \eqref{eq:new-kl-R} can be written as
\begin{equation}\label{eq:new-kl-per-step}
   D_{\mathrm{KL}}(P_{\to} (\mathbf{y} ) \|  P_{\leftarrow} (J\mathbf{y}))
  = \frac{1}{2}\sum_{s=1}^{T}
  \Bigl[
      \operatorname{tr} \bigl(S^{-1} \Sigma_s^B\bigr) - n_y
  \Bigr].
\end{equation}

By noticing that
\begin{equation}
\mathbb{E}_{P_{\to}} \Bigl[
     (e_s^B)^\top S^{-1}  e_s^B
  \Bigr] = \operatorname{tr} \bigl(S^{-1} \Sigma_s^B\bigr),
\end{equation}
\eqref{eq:new-kl-per-step} can be further rewritten as
\begin{equation}
\begin{split}
  &D_{\mathrm{KL}}(P_{\to} (\mathbf{y} ) \|  P_{\leftarrow} (J\mathbf{y})) \\
 &{} \quad  = \frac{1}{2} \sum_{s=1}^{T}  \left( \mathbb{E}_{P_{\to}} \Bigl[
    (e_s^B)^\top S^{-1}  e_s^B
  \Bigr] - n_y \right).
  \end{split}
  \label{eq:kl_innov}
\end{equation}
 
 As a consistency check, we directly calculate the log-ratio of the forward and backward path probabilities from \eqref{eq:pF} and \eqref{eq:pR}  (i.e., the stochastic entropy production):
 \begin{equation}\label{linear_log_ratio}
\sigma (\mathbf{y}) = \ln \frac{P_{\to} (\mathbf{y} )}{P_{\leftarrow} (J\mathbf{y})} =
\frac{1}{2}\sum_{s=1}^{T} ( e_s^B)^\top S^{-1}  e_s^B
-\frac{1}{2}\sum_{t=1}^{T} e_t^\top S^{-1} e_t,
 \end{equation}
 where we used $\tilde{e}_s^B = e_s^B$ for the realization $\tilde{y}_s = y_{T-s+1}$ (see the discussion above).
 Indeed, \eqref{eq:kl_innov} is reproduced from
 \begin{equation}
 \mathbb{E}_{P_{\to}}\Biggl[ \ln \frac{P_{\to} (\mathbf{y} )}{P_{\leftarrow} (J\mathbf{y})} \Biggr] =  \frac{1}{2} \sum_{s=1}^{T}  \left( \mathbb{E}_{P_{\to}} \Bigl[
    (e_s^B)^\top S^{-1}  e_s^B
  \Bigr] - n_y \right),
 \end{equation}
 where we used
 \begin{equation}
 \mathbb{E}_{P_{\to}}[ e_t^\top S^{-1} e_t] =   \mathbb{E}_{P_{\to}} \Bigl[  \operatorname{tr}[ S^{-1} e_t e_t^\top ] \Bigr] = \operatorname{tr} [I_{n_y}] = n_y.
 \end{equation}

The expression~\eqref{eq:kl_innov} provides an operational meaning of the entropy production.
Under the forward measure~$P_{\to}$, the covariance
$\Sigma_s^B$ of the backward innovation $e_s^B$ generally differs
from~$S$, and~\eqref{eq:kl_innov} quantifies this cumulative
mismatch between the backward innovation statistics expected by the
model and those actually realized under the forward dynamics.

Finally, we remark on the asymptotic regime $T \to \infty$, where we assume that $A$ is stable (that is, all eigenvalues strictly lie inside the unit circle).
In the scalar case ($n_y = 1$), every stationary Gaussian process
is time-reversible~\cite{Weiss1975}, so the entropy production
remains bounded as $T \to \infty$ and is a boundary
effect of the deterministic initial condition
$\hat{x}_{1|0} = 0$.
In the multivariate case ($n_y > 1$), stationary Gaussian
processes can be time-irreversible whenever the
cross-covariance matrices are not
symmetric~\cite{TongZhang2005,GeorgiouLindquist2014},
and the entropy production can grow linearly with~$T$.

\noindent\textbf{Scalar case ($n_x = n_y = 1$).} 
When $n_x = n_y = 1$, the factors $S^{-1}$ and $S$
in~\eqref{eq:new-kl-R} cancel, yielding
\begin{equation}\label{eq:new-kl-scalar}
  D_{\mathrm{KL}}(P_{\to} (\mathbf{y} ) \|  P_{\leftarrow} (J\mathbf{y}))
  = \frac{1}{2}\bigl(\|\mathcal{R}\|_F^2 - T\bigr),
\end{equation}
where $\|\mathcal{R}\|_F^2 = \sum_{s,k} \mathcal{R}_{sk}^2$
is the squared Frobenius norm.
 
We consider a further specific case: $T = 2$.
Let $H \equiv CAK = H_1$.
Then
\begin{equation}
  \mathcal{H}
  = \begin{pmatrix} 1 & 0 \\ H & 1 \end{pmatrix},
  \quad
  \mathcal{G}
  = \begin{pmatrix} 1 & 0 \\ -H & 1 \end{pmatrix},
  \quad
  J
  = \begin{pmatrix} 0 & 1 \\ 1 & 0 \end{pmatrix},
\end{equation}
so that
\begin{equation}\label{eq:new-R-T2}
  \mathcal{R}
  = \mathcal{G} J \mathcal{H}
  = \begin{pmatrix} H & 1 \\ 1-H^2 & -H \end{pmatrix}.
\end{equation}
Hence
$\|\mathcal{R}\|_F^2 =  2 + H^4$,
and
\begin{equation}\label{eq:new-kl-T2}
   D_{\mathrm{KL}}(P_{\to} (\mathbf{y} ) \|  P_{\leftarrow} (J\mathbf{y})) = \frac{H^4}{2}
  = \frac{(CAK)^4}{2}.
\end{equation}

%----------------------------------------------------------------------
\subsection{Numerical verification by Monte Carlo sampling}
\label{subsec:numerical-verification}
%----------------------------------------------------------------------

\begin{figure}[htbp]
  \centering
  \includegraphics[width=1.0\linewidth]{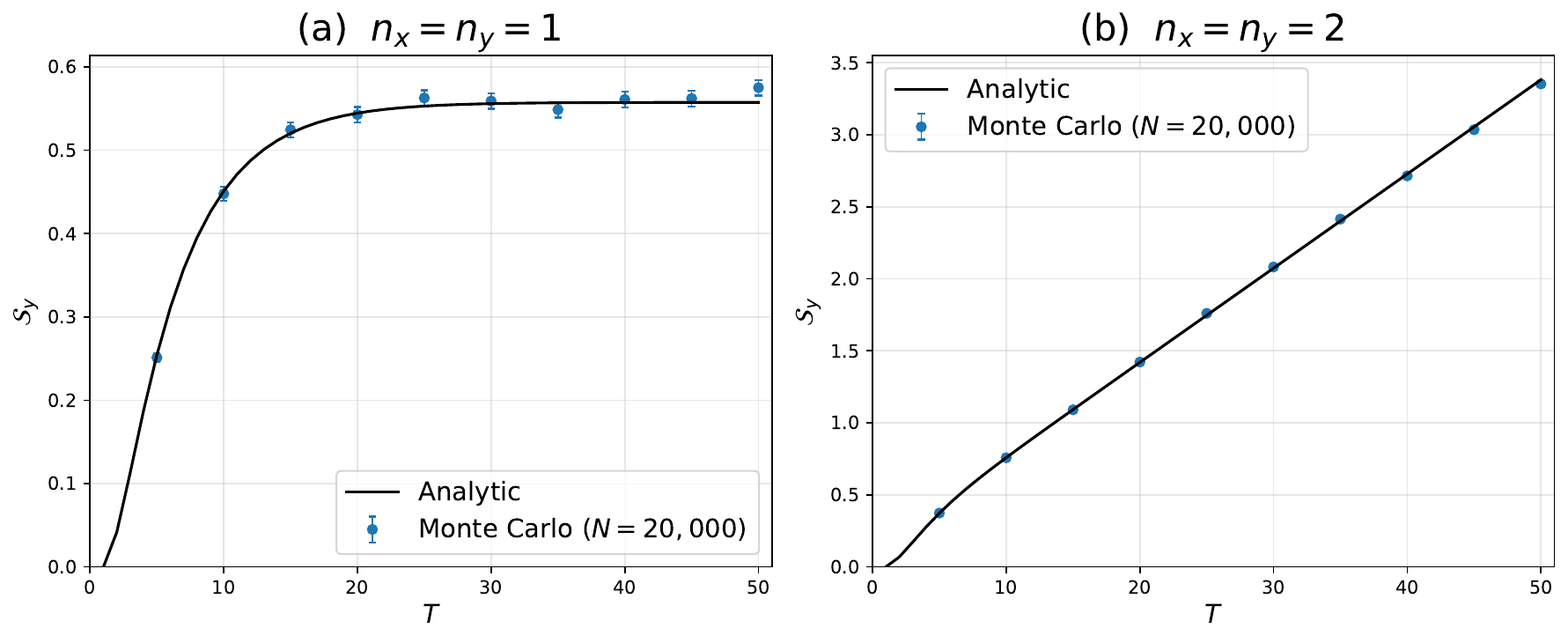}
  \caption{Numerical verification of the analytical entropy
  production~\eqref{eq:new-kl-R}
  by Monte Carlo sampling~\eqref{eq:sigma-per-traj}--\eqref{eq:sigma-MC}
  with $N = 20{,}000$ trajectories.
  Solid curves: analytical values;
  circles with error bars: Monte Carlo estimates.
  Error bars indicate $\pm 1$ standard error of the mean,
  $\mathrm{SE} = \mathrm{std}(\sigma) / \sqrt{N}$.
  (a)~Scalar case ($n_x = n_y = 1$) with
  $A = 0.9$, $C = 1$, $Q = 1$, $R = 1$, and
  (b)~Multivariate case ($n_x = n_y = 2$) with
  $A = \bigl(\begin{smallmatrix} 0.8 & 0.3 \\ 0 & 0.5 \end{smallmatrix}\bigr)$,
  $C = \bigl(\begin{smallmatrix} 1 & 0.5 \\ 0 & 1 \end{smallmatrix}\bigr)$,
  $Q = I_2$, $R = I_2$.}
  \label{fig:numerical-verification}
\end{figure}

As a demonstration of the sampling procedure proposed in
Section~\ref{subsec:cost-examples},
we numerically verify the analytical expression~\eqref{eq:new-kl-R}
for the entropy production
in the linear Gaussian setting.
Two parameter sets for~\eqref{eq:state}--\eqref{eq:obs} are examined:
a scalar case ($n_x = n_y = 1$) and a multivariate case
($n_x = n_y = 2$); the specific values are listed in the
caption of Figure~\ref{fig:numerical-verification}.
All eigenvalues of~$A$ lie strictly inside the unit circle.
We set $\hat{x}_{1|0} = \hat{x}_{1|0}^B = 0$.

For each sequence length $T$, $N = 20{,}000$ trajectories
$y_{1:T}^{(n)}$ are sampled from the forward generative
process (Section~\ref{sec:gen}), and
the stochastic entropy
production~$\sigma(y_{1:T}^{(n)})$~\eqref{eq:sigma-per-traj}
is computed via the forward and backward passes
as described in Section~\ref{subsec:cost-examples}.
In the present Gaussian setting, the normalization constants
of the emission kernels cancel between the forward and backward sums,
reducing $\sigma$ to the difference in quadratic forms as shown in \eqref{linear_log_ratio}.
The entropy production is then estimated
by~\eqref{eq:sigma-MC}.

Figure~\ref{fig:numerical-verification} shows the analytical
entropy production~\eqref{eq:new-kl-R} together with
the Monte Carlo estimates
as functions of~$T$.
For both parameter sets, the Monte Carlo estimates are
in good agreement with the analytical values
across the entire range $T = 5, 10, \dots, 50$.
In the scalar case~(a), the entropy production saturates
to a finite value, consistent with the time-reversibility of
stationary scalar Gaussian processes~\cite{Weiss1975}.
In the multivariate case~(b), the entropy production grows
approximately linearly with~$T$,
reflecting the genuine time-irreversibility of the multivariate
process~\cite{TongZhang2005,GeorgiouLindquist2014}.

 We emphasize that the Monte Carlo procedure involves stochastic
sampling only in the forward direction.
For each sample trajectory $y_{1:T}^{(n)} \sim P_{\to}$,
the forward innovations $e_t$ are drawn independently from
$\mathcal{N}(0, S)$, from which $y_{1:T}$ is generated
via~\eqref{eq:gen_y}--\eqref{eq:predict}.
The backward innovations $e_s^B$ are then obtained
deterministically from the same $y_{1:T}$
via~\eqref{eq:eB_extract}--\eqref{eq:predictB},
without any additional random sampling.
As discussed in Section~\ref{sec:rev}, for the realization
$\tilde{y}_s = y_{T-s+1}$, the quantities $e_s^B$ and
$\tilde{e}_s^B$ take the same numerical value, but their
statistical properties under $P_{\to}$ differ from the i.i.d.\
$\mathcal{N}(0, S)$ assumed by the backward model.

%======================================================================
\section{Retrospective decompositions of entropy production}
\label{sec:retro}
%======================================================================

Turning to the general setup of Section~\ref{sec:setup}, we derive exact decompositions of the entropy production that hold in the general, non-recursive setting.
The entropy production is first split into non-negative per-step contributions via retrospective Bayesian inference, and each contribution is further decomposed into a compression loss and a model mismatch.
These decompositions reveal a set of information-theoretic structures and are of fundamental interest for the thermodynamics of information in non-Markovian processes.
Note that  \emph{retrospective} refers only to the Bayesian decomposition below, not to the definition of the  protocol-reversed backward process introduced in Section~\ref{sec:backward_ep}. 

\subsection{Per-step decomposition}

We decompose the entropy production into a sum of non-negative per-step terms by comparing the forward and backward path measures at each time step through retrospective Bayesian inference.

From the standard Bayesian rule (i.e., the chain rule),
\begin{equation}\label{retroBayesian}
P_{\to}(y_{t:T}) = P_{\to}(y_t \mid y_{t+1:T}) P_{\to}(y_{t+1:T}).
\end{equation}
By applying this iteratively, we obtain
\begin{equation}
P_{\to}(y_{1:T} )= \prod_{t=1}^{T} P_{\to}(y_t \mid y_{t+1:T}).
\end{equation}
Also by definition
\begin{equation}
P_{\leftarrow}(y_{T:1}) = \prod_{t=1}^T p_t(y_t \mid g_{t+1}^{\leftarrow}(y_{T:t+1})).
\end{equation}
Thus,
\begin{equation}
\begin{split}
  \mathcal{S}_y
   &=  \mathbb{E}_{P_{\to}} \left[
    \ln \frac{P_{\to}(y_{1:T})}{P_{\leftarrow}(y_{T:1})}
  \right] \\
    & = 
   \mathbb{E}_{P_{\to}} \Big[
   \sum_{t=1}^{T} \ln \frac{P_{\to}(y_t \mid y_{t+1:T})}{p_t(y_t \mid g_{t+1}^{\leftarrow}(y_{T:t+1}))}
   \Big].
   \end{split}
   \end{equation}
  We finally obtain
  \begin{equation}\label{eq:Sigma-exact}
  \mathcal{S}_y
  = \sum_{t=1}^{T} \mathcal{D}_t,
\end{equation}
where
$  \mathcal{D}_t 
 \equiv 
  \mathbb{E}_{y_{t+1:T}\sim P_{\to}} \Big[
    D_{\mathrm{KL}} \bigl(
      P_{\to}(y_t \mid y_{t+1:T})
       \big\| 
      p_t(y_t \mid g_{t+1}^{\leftarrow}(y_{T:t+1}))
    \bigr)
  \Big] 
   \geq  0$,
or equivalently,
\begin{equation}\label{eq:D-t-logratio}
  \mathcal{D}_t
  \equiv
  \mathbb{E}_{P_{\to}} \left[
    \ln \frac{P_{\to}(y_t \mid y_{t+1:T})}
             {p_t \bigl(y_t \mid g_{t+1}^{\leftarrow}(y_{T:t+1})\bigr)}
  \right].
\end{equation}

The above result implies that the total entropy production can be decomposed into non-negative terms associated with individual time steps, even when the entire process is non-Markovian.
From this, the equality $ \mathcal{S}_y = 0$  holds if and only if $\mathcal{D}_t = 0$ for every $t$,
equivalently, if and only if
$  p_t \bigl(\cdot \mid g_{t+1}^{\leftarrow}(y_{T:t+1})\bigr)
  = P_{\to} \bigl(\cdot \mid y_{t+1:T}\bigr)$
for almost every $y_{t+1:T}$ and every $t$.

The quantity $\mathcal{D}_t$ is the expected mismatch between
the Bayesian retrospective distribution $P_{\to}(y_t \mid y_{t+1:T})$
and the conditional distribution used by the actual backward model.
This sheds light on the concept of entropy production; the total entropy production can be interpreted as the gap between the retrospective Bayesian distribution and the physical (that is, protocol-reversed) backward model.
In the spirit of variational inference theory \cite{KingmaWelling2014,HoffmanJohnson2016,AlemiFischerDillonMurphy2018}, $p_t \bigl(y_t \mid g_{t+1}^{\leftarrow}(y_{T:t+1})\bigr)$ can be interpreted as a test function that approximates the Bayesian posterior $P_{\to}(y_t \mid y_{t+1:T})$.

We note that the reverse process employed in diffusion models~\cite{SohlDickstein2015,Ho2020,Song2021} corresponds to the Bayesian retrodiction, where the time evolution of the probability distribution itself is time-reversed~\cite{Anderson1982}.
This is conceptually distinct from the backward process in stochastic thermodynamics, where the control protocol is time-reversed.
Their gap is measured precisely by the entropy production for continuous-time diffusion models~\cite{Premkumar2025}; our result above represents another manifestation of this gap.

A simple everyday example may help illustrate this gap.
Consider a statement: ``If I don't study, then my mom will get angry.''
Here, we take $y_1$ as an episode ``I don't study'' and $y_2$ as ``my mom gets angry.''
The retrospective Bayesian inference is ``If my mom gets angry, that implies I didn't study,'' which we intuitively expect to be true in daily life (its probability should be close to one). (Note that the contrapositive is ``If my mom doesn't get angry, that implies I studied,'' which is logically true if the original statement is true.)
On the other hand, the backward process of our framework is given by reversing the real-time ordering, ``If my mom gets angry, then I will not study,'' which we intuitively expect to be false  in daily life (its probability should be close to zero).
Therefore, this daily-life situation is highly irreversible.

\subsection{Compression loss and model mismatch}

We further decompose each per-step contribution $\mathcal{D}_t$ into two separately non-negative terms: a compression loss, which quantifies the retrospective information discarded by the backward summary, and a model mismatch, which quantifies the cost of reusing the forward emission kernel in the backward direction.

Write $g^{\leftarrow}_{t+1} \equiv g_{t+1}^{\leftarrow}(y_{T:t+1})$ for brevity.
For each fixed $y_{t+1:T}$, insert the intermediate distribution
$P_{\to}(y_t \mid g^{\leftarrow}_{t+1})$ into the KL divergence:
\begin{equation}
\begin{split}
  &D_{\mathrm{KL}} \bigl(
    P_{\to}(y_t \mid y_{t+1:T})
     \big\| 
    p_t(y_t \mid g^{\leftarrow}_{t+1})
  \bigr) \\
  &\quad =
  D_{\mathrm{KL}} \bigl(
    P_{\to}(y_t \mid y_{t+1:T})
     \big\| 
    P_{\to}(y_t \mid g^{\leftarrow}_{t+1})
  \bigr) \\
  &\quad \quad   + 
  \sum_{y_t} P_{\to}(y_t \mid y_{t+1:T})
    \ln \frac{P_{\to}(y_t \mid g^{\leftarrow}_{t+1})}{p_t(y_t \mid g^{\leftarrow}_{t+1})}.
    \end{split}
  \label{eq:KL-split}
\end{equation}
Here, $P_{\to}(y_t \mid g^{\leftarrow}_{t+1})$ denotes the Bayesian retrospective distribution of $y_t$ conditioned on  $g^{\leftarrow}_{t+1}$, instead of the full future $y_{t+1:T}$.
Taking the expectation over $y_{t+1:T} \sim P_{\to}$, the two terms become the terms called the compression loss and the model mismatch as follows.

\noindent\textbf{Compression loss.}
Since $g^{\leftarrow}_{t+1}$ is a deterministic
function of $y_{t+1:T}$, $ P_{\to}(y_t \mid y_{t+1:T}) =  P_{\to}(y_t \mid y_{t+1:T}, g^{\leftarrow}_{t+1} )$. Therefore,
\begin{equation}\label{eq:compression-loss}
\begin{split}
  &\mathbb{E}_{y_{t+1:T}} \Big[
    D_{\mathrm{KL}} \bigl(
      P_{\to}(y_t \mid y_{t+1:T})
       \big\| 
      P_{\to}(y_t \mid g^{\leftarrow}_{t+1})
    \bigr)
  \Big] \\
  &\quad = I_{P_{\to}} \bigl(y_t;  y_{t+1:T} \mid g^{\leftarrow}_{t+1}\bigr) \\
  &\quad \equiv
   \mathcal{L}_t,
   \end{split}
\end{equation}
where $ I_{P_{\to}} \bigl(y_t;  y_{t+1:T} \mid g^{\leftarrow}_{t+1}\bigr)$ is the mutual information between $y_t$ and $y_{t+1:T}$  under the condition of $ g^{\leftarrow}_{t+1}$, with respect to the forward distribution $P_\to$.
This is the information about $y_t$ contained in the full future
$y_{t+1:T}$ that is discarded by the compression from $y_{t+1:T}$ to $g^{\leftarrow}_{t+1}$.

\noindent\textbf{Model mismatch.}
The logarithmic factor in the second term depends on $y_{t+1:T}$
only through $g^{\leftarrow}_{t+1}$.
Hence
\begin{align}
  &\sum_{y_{t+1:T}} P_{\to}(y_{t+1:T})
    \sum_{y_t} P_{\to}(y_t \mid y_{t+1:T}) 
    \ln \frac{P_{\to}(y_t \mid g^{\leftarrow}_{t+1})}{p_t(y_t \mid g^{\leftarrow}_{t+1})}
  \notag \\
  &\quad =
  \sum_{g^{\leftarrow}_{t+1}} P_{\to}(g^{\leftarrow}_{t+1})
    \sum_{y_t} P_{\to}(y_t \mid g^{\leftarrow}_{t+1}) 
    \ln \frac{P_{\to}(y_t \mid g^{\leftarrow}_{t+1})}{p_t(y_t \mid g^{\leftarrow}_{t+1})}.
  \label{eq:mismatch-marginalise}
\end{align}
Therefore the second term in \eqref{eq:KL-split} equals
\begin{equation}\label{eq:model-mismatch}
  \mathbb{E}_{g^{\leftarrow}_{t+1} \sim P_{\to}} \Big[
    D_{\mathrm{KL}} \bigl(
      P_{\to}(y_t \mid g^{\leftarrow}_{t+1})
       \big\| 
      p_t(y_t \mid g^{\leftarrow}_{t+1})
    \bigr)
  \Big]
  \equiv
  \mathcal{M}_t.
\end{equation}
This represents  the additional cost of using
    $p_t(\cdot \mid g^{\leftarrow}_{t+1})$ instead of the Bayesian retrospective
    distribution $P_{\to}(\cdot \mid g^{\leftarrow}_{t+1})$.

\noindent\textbf{The decomposition.}
Combining \eqref{eq:compression-loss} and \eqref{eq:model-mismatch} gives
\begin{equation}\label{eq:D-t-refined}
  \mathcal{D}_t
  = \mathcal{L}_t + \mathcal{M}_t,
\end{equation}
where  $\mathcal{L}_t$ and $\mathcal{M}_t$ are non-negative.

The decomposition \eqref{eq:D-t-refined} is formally similar to the
ELBO decompositions familiar from variational inference
\cite{KingmaWelling2014,HoffmanJohnson2016,AlemiFischerDillonMurphy2018},
where the average KL term in the ELBO
likewise splits into a mutual-information term and a distributional-mismatch
term.
The shared algebraic origin is the chain rule for KL divergence applied
via an intermediate distribution.
In the present setting, however, the object being decomposed is not  the KL term of an ELBO
but a per-step contribution to the entropy production
$\mathcal{S}_y$, defined through the path-probability ratio
\eqref{eq:Sigma-def}.
Accordingly, $\mathcal{L}_t$ \eqref{eq:compression-loss} and
$\mathcal{M}_t$ \eqref{eq:model-mismatch} acquire a physical
interpretation as distinct sources of temporal irreversibility; the
former from lossy compression of the future, the latter from reuse of
the forward emission kernel.
We note that the term of mismatch cost also appears 
in the stochastic thermodynamics of 
computation~\cite{WolpertKolchinsky2020,ManzanoKardesRoldanWolpert2024}
with a related but distinct meaning, quantifying the 
extra dissipation due to a non-optimal initial distribution 
of a physical process.

\subsection{Refined second law}

Combining the compression-loss decomposition with the chain rule for mutual information, we obtain a lower bound on the entropy production in terms of the gap between the predictive information carried by the forward and backward latent-state summaries.

Since $g_{t+1}^{\leftarrow}$ is a deterministic function of
$y_{t+1:T}$, the chain rule for mutual information gives
\begin{equation}\label{eq:L-as-MI-diff}
  \mathcal{L}_t
  = I_{P_{\to}}(y_t;  y_{t+1:T})
    - I_{P_{\to}} \bigl(
      y_t;  g_{t+1}^{\leftarrow}
    \bigr).
\end{equation}
On the forward side,
since $f_{t-1}^{\to}(y_{1:t-1})$ is a sufficient summary of the past for predicting
$y_t$, 
\begin{equation}
  I_{P_{\to}}(y_t;  y_{1:t-1})
  = I_{P_{\to}} \bigl(y_t;  f_{t-1}^{\to}(y_{1:t-1})\bigr).
\end{equation}
That is, the forward summary loses no predictive information by construction of the dynamics, whereas the backward
summary generally does.
We also note that the forward and reverse chain rules for conditional Shannon entropies, written as $H_{P_\to} (\cdot \mid \cdot)$, give
\begin{equation}\label{eq:entropy-chain-cancel}
  \sum_{t=1}^{T} H_{P_{\to}}(y_t \mid y_{1:t-1})
  = H_{P_{\to}}(y_{1:T})
  = \sum_{t=1}^{T} H_{P_{\to}}(y_t \mid y_{t+1:T}),
\end{equation}
and thus
\begin{equation}
\sum_{t=1}^T I_{P_{\to}}(y_t;  y_{1:t-1})
 = \sum_{t=1}^T I_{P_{\to}}(y_t;  y_{t+1:T}).
\end{equation}
Combining all of the above, we obtain
\begin{equation}
  \sum_{t=1}^T \mathcal{L}_t
  =
  \sum_{t=1}^{T}
    \bigl[
      I_{P_{\to}} \bigl(y_t;  f_{t-1}^{\to}(y_{1:t-1})\bigr)
      -
      I_{P_{\to}} \bigl(
        y_t;  g_{t+1}^{\leftarrow}(y_{T:t+1})
      \bigr)
    \bigr],
\end{equation}
where the raw past and future are replaced by their deterministic summaries
$f_{t-1}^{\to}$ and $g_{t+1}^{\leftarrow}$.
Therefore,
\begin{equation}\label{eq:refined-latent}
\begin{split}
  \mathcal{S}_y
   &\geq 
  \sum_{t=1}^{T}
    \bigl[
      I_{P_{\to}} \bigl(y_t;  f_{t-1}^{\to}(y_{1:t-1})\bigr)
      -
      I_{P_{\to}} \bigl(
        y_t;  g_{t+1}^{\leftarrow}(y_{T:t+1})
      \bigr)
    \bigr] \\
    &\geq 0.
    \end{split}
\end{equation}
This is regarded as a  refined version of the second law; the entropy production is bounded from below by the gap between the mutual information terms associated with the past and the future.
The more memory of the past the current state has lost, or the more the future summary retains of the information about the current state, the smaller the entropy production can be.
In the special case where $g^{\leftarrow}_{t+1}$ is injective for all $t$, every $\mathcal{L}_t$ vanishes, so that $\mathcal{D}_t = \mathcal{M}_t$ and inequality (99) reduces to $\mathcal{S}_y \geq 0$.

We note that  Ref.~\cite{StillSivakBellCrooks2012} related dissipation to the
predictive information retained by a driven system, but assumed  that the
system is Markovian and did not define the entropy production at the level of an observed
non-Markovian output.
We emphasize that their main result in \cite{StillSivakBellCrooks2012} is distinct from our result \eqref{eq:refined-latent} even in the Markovian case.

Another structurally relevant concept is the \emph{backward transfer
entropy} (BTE)~\cite{Ito2016BTE}, which considers a bipartite stochastic process $(x_t, y_t)$ and defines
the backward transfer entropy as the transfer entropy computed on
time-reversed trajectories.
Expressed in forward-time variables, the BTE between two variables $x,y$ is defined as
\begin{equation}
  T^B_{y \to x}(t)
  = I \bigl(y_{t+1:T};  x_t \mid x_{t+1:T}\bigr),
\end{equation}
which quantifies how much the future of $y$ retrodicts the present
of $x$ beyond what the future of $x$ itself reveals.
The compression-loss term of the present work, $\mathcal{L}_t$ in \eqref{eq:compression-loss},
shares a similar information-theoretic form.
However, the BTE is an inter-variable quantity: it measures the
retrospective information that one subsystem ($y$) carries about a
 distinct subsystem ($x$).
In contrast, $\mathcal{L}_t$ is an intra-variable quantity
involving a single observed process $y_t$ at different times;
the conditioning variable $g_{t+1}^{\leftarrow}(y_{T:t+1})$ is a
deterministic function of $y$'s own future.

%----------------------------------------------------------------------
\subsection{Estimation of retrospective decomposition terms}
\label{subsec:not-estimable}
%----------------------------------------------------------------------

While $\mathcal{S}_y$ is directly computable by Monte Carlo sampling, the
individual decomposition terms $\mathcal{D}_t$, $\mathcal{L}_t$,
and $\mathcal{M}_t$ are not.
The obstacle is the Bayesian retrospective distribution
$P_{\to}(y_t \mid y_{t+1:T})$, which appears in the definition
of $\mathcal{D}_t$ \eqref{eq:D-t-logratio}.
This distribution is obtained by marginalizing the forward path
measure over all possible pasts:
\begin{equation}\label{eq:retrospective-marginal}
  P_{\to}(y_t \mid y_{t+1:T})
  = \frac{%
   \sum_{y_1, \dots, y_{t-1}} P_{\to}(y_{1:T})
  }{%
   \sum_{y_1, \dots, y_{t}}  P_{\to}(y_{1:T})
  }.
\end{equation}
Even though each factor $p_t(y_{t+1} \mid h_{t})$ in the forward path
probability is explicitly evaluable, the summations in the numerator
and denominator couple all time steps through the deterministic maps
$\Phi_t$.
Unlike the forward conditional
$P_{\to}(y_{t+1} \mid y_{1:t}) = p_t(y_{t+1} \mid h_{t})$, which the
model provides by construction, the retrospective conditional
$P_{\to}(y_t \mid y_{t+1:T})$ is not directly supplied by any
component of the autoregressive architecture.

As discussed above, the Bayesian retrospective distribution
$P_{\to}(y_t \mid y_{t+1:T})$ plays a role analogous to the
reverse-time transition in diffusion
models~\cite{SohlDickstein2015,Ho2020,Song2021,Anderson1982},
where  the intractable reverse-time transition is
approximated by a neural network trained on samples from the
forward process.
This suggests a parallel strategy for the present setting:
training a reverse-direction autoregressive model
(e.g., $r_{\theta}(y_t | y_{t+1:T})$ with model parameters $\theta$) on trajectories sampled from the
forward process to approximate $P_{\to}(y_t \mid y_{t+1:T})$,
from which the individual terms $\mathcal{D}_t$, $\mathcal{L}_t$,
and $\mathcal{M}_t$ could be estimated.
Developing such estimation procedures remains an open problem.

%======================================================================
\section{Summary and perspectives}
\label{sec:summary}
%======================================================================

We have developed a stochastic-thermodynamic framework for non-Markovian processes generated by autoregressive models with deterministic internal memory, encompassing Transformers, RNNs, the Kalman filter, state space models, and Mamba under a single formalism (Table~\ref{tab:examples}). The entropy production $\mathcal{S}_y$, defined as the KL divergence between the forward and backward path measures~\eqref{eq:Sigma-def}, is efficiently computable from sampled trajectories without combinatorial overhead from enumerating histories, owing to the deterministic latent state and the explicit emission kernel (Section~\ref{sec:sampling}); the same cost applies to the temporally coarse-grained variant that reverses blocks rather than individual tokens. We have demonstrated the framework through a proof-of-concept experiment on GPT-2 (Section~\ref{sec:gpt2-demo}), 
where the sentence-level entropy production 
tends to be larger for causally ordered than for non-causal texts (with statistically significant differences for three of the five corpora with GPT-2, and for all five with a larger evaluator,
Qwen3-4B-Base; Appendix~\ref{app:qwen3-fixed-text}).
Moreover, we have analyzed the linear Gaussian (Kalman) case (Section~\ref{sec:linear-gaussian-example}), where the analytical entropy production is confirmed by Monte Carlo sampling (Figure~\ref{fig:numerical-verification}). We have further shown that $\mathcal{S}_y$ decomposes exactly into non-negative per-step contributions $\mathcal{D}_t$~\eqref{eq:Sigma-exact}, each splitting into a compression loss $\mathcal{L}_t$ and a model mismatch $\mathcal{M}_t$~\eqref{eq:D-t-refined}.
This decomposition is formally reminiscent of the ELBO decompositions for variational inference~\cite{KingmaWelling2014,HoffmanJohnson2016,AlemiFischerDillonMurphy2018}, which reveals an interesting connection between fundamental concepts in stochastic thermodynamics and machine learning.

From the viewpoint of stochastic thermodynamics, an interesting direction concerns finite-time trade-off
relations.
In Markovian stochastic thermodynamics, thermodynamic uncertainty
relations~\cite{BaratoSeifert2015,HorowitzGingrich2020} and
thermodynamic speed
limits~\cite{ShiraishiFunoSaito2018} constrain the interplay among
the precision of currents, the speed of state transformations, and
the entropy production.
Indeed, such trade-off relations have been demonstrated for diffusion models~\cite{IkedaUdaOkanoharaIto2025} and dense associative memory networks~\cite{RookeKrotovBalasubramanianWolpert2026}.
Extending these relations to the non-Markovian
autoregressive setting could yield new bounds linking
the speed, accuracy, and irreversibility of sequence generation in
autoregressive generative models.

In standard implementations, LLMs often terminate generation 
upon producing an end-of-sequence (EOS) token, making the 
sequence length a stopping time.
Refs.~\cite{Manzano2021,ManzanoKardesRoldanWolpert2024} developed 
thermodynamic bounds at stochastic stopping times, 
which could be applied to generalize the present framework to
variable-length processes.
It would also be of interest to explore connections with computational mechanics \cite{ShaliziCrutchfield2001,BoydMandalCrutchfield2018,BoydMandalCrutchfield2017}, where the minimal sufficient statistic of the past (the causal state or $\epsilon$-machine) plays a role analogous to the forward latent state. 

Applying our framework to
larger and more capable language models than those examined here remains an important
direction for future research and raises further challenges
at different levels.
At the technical level, a single semantic episode can often be
expressed by multiple distinct token sequences
(e.g., paraphrases),
so that coarse-graining at the token-sequence level may not
fully capture the irreversibility at the level of meaning;
an additional coarse-graining over token sequences that encode the
same semantic content may be needed.
At a more fundamental level, the block-level entropy production
reflects at least three intertwined contributions:
genuine causal dependence among the described events,
mere temporal ordering without causal relationship,
and the conventions of discourse structure
(e.g., narrative arc, rhetorical organization).
Disentangling these contributions remains
a highly nontrivial open problem.
If these challenges can be addressed, the entropy
production could provide a quantitative probe of
irreversibility and causality of the
real-world processes whose structure is implicitly encoded in the
internal representations of an LLM --- often referred to as
a world model in the machine-learning
literature.

\section*{Acknowledgments}
Claude Opus 4.6, Claude Fable 5, GPT-5.4 Thinking/Pro, and GPT-5.6 Sol were used for assistance with manuscript preparation. The author takes full responsibility for the contents. 

The author is  grateful to Daisuke Okanohara, Ken Funo, Igen Koh, and David H. Wolpert for valuable discussions. 
The author is also grateful to Shoki Sugimoto for assistance with code review.
 This work was supported by JST ERATO Grant No. JPMJER2302, Japan, and also by Institute for AI and Beyond of the University of Tokyo.
 
\section*{Data Availability}
The code and data used in this study are available at Ref.~\cite{code}.

\appendix

%======================================================================
\section{Markovian embedding}
\label{sec:markovian-embedding}
%======================================================================

In this appendix, we clarify the relationship between the autoregressive framework developed in the main text and the concept of Markovian embedding.

\subsection{General definition}

In this paper, we say that a process $\mathbf{x}_t$ provides a
\emph{Markovian embedding} of the observed non-Markovian process $y_t$ if
$\mathbf{x}_t$ is Markovian and the joint law factorizes as
\begin{equation}\label{eq:markov-embedding-def}
  P_{\to}(\mathbf{x}_{1:T},  y_{1:T})
  =
  p(\mathbf{x}_1)
  \prod_{t=1}^{T-1} p_t(\mathbf{x}_{t+1} \mid \mathbf{x}_t)
  \prod_{t=1}^{T} q_t(y_t \mid \mathbf{x}_t).
\end{equation}
That is, $y_t$ is emitted memorylessly from the Markovian state
$\mathbf{x}_t$ at each time step.
This definition can be graphically represented as follows, where $\Rightarrow$ and $\Downarrow$ describe stochastic influences.
\begin{equation}\label{eq:graphical2}
\begin{matrix}
\cdots & \mathbf{x}_{t-2} & \Rightarrow & \mathbf{x}_{t-1} &\Rightarrow & \mathbf{x}_t &\Rightarrow & \mathbf{x}_{t+1} & \cdots \\
       & \Downarrow &   & \Downarrow &       & \Downarrow &     & \Downarrow &        \\
\cdots & y_{t-2} &                 & y_{t-1} &                 & y_t      &              & y_{t+1}  & \cdots
\end{matrix}
\end{equation}
Note that a more standard definition of Markovian embedding is a special case of the above definition, where $y_t$ is obtained from $\mathbf{x}_t$ by a deterministic projection~\cite{KemenySnell1960}.

The corresponding backward process is defined as
\begin{equation}\label{eq:backward-ME}
  P_{\leftarrow}(\tilde{\mathbf{x}}_{1:T},  \tilde y_{1:T})
  =
  \tilde{p}(\tilde{\mathbf{x}}_1)
  \prod_{s=1}^{T-1} \tilde p_s(\tilde{\mathbf{x}}_{s+1} \mid \tilde{\mathbf{x}}_{s})
 \prod_{s=1}^{T} \tilde q_{s}(\tilde y_s \mid \tilde{\mathbf{x}}_s),
\end{equation}
where we choose $\tilde p_s = p_{T-s}$ and $\tilde{q}_s = q_{T-s+1}$ to be consistent with Crooks' notion of the backward process~\cite{Crooks1999}.
Here, we assume that the backward $y$-marginal coincides with the distribution defined in Section~\ref{sec:backward}, but we do not assume \eqref{eq:assume_final}.

We define the $\mathbf{x}$-marginal and $y$-marginal of these path distributions, and write  $P_{\to}(\mathbf{x}_{1:T})$, $P_{\leftarrow}(\mathbf{x}_{T:1})$, $P_{\to}(y_{1:T})$, $P_{\leftarrow}(y_{T:1})$ by substituting $\tilde{\mathbf{x}}_{1:T} = \mathbf{x}_{T:1}$ and $\tilde{y}_{1:T} = y_{T:1}$ for the backward process.
 We  then introduce the entropy production associated with $\mathbf{x}$
\begin{equation}\label{eq:Sigma-x-def}
  \mathcal{S}_{\mathbf{x}}
   \equiv 
  D_{\mathrm{KL}} \bigl(
    P_{\to}(\mathbf{x}_{1:T})
     \big\| 
    P_{\leftarrow}(\mathbf{x}_{T:1})
  \bigr),
\end{equation}
and with $y$
\begin{equation}\label{eq:Sigma-y-ME-def}
  \mathcal{S}_y
   \equiv
  D_{\mathrm{KL}} \bigl(
    P_{\to}(y_{1:T})
     \big\| 
     P_{\leftarrow}(y_{T:1})
  \bigr).
\end{equation}

Here, the product channel
$\mathbf{x}_{1:T} \mapsto y_{1:T}$ defined by
$\prod_{t=1}^{T} q_t (y_t \mid \mathbf{x}_t) = \prod_{s=1}^{T} \tilde q_{s}(y_{T-s+1} \mid \mathbf{x}_{T-s+1})$
is a stochastic map (``Markovian'' map) from $\mathbf{x}$-paths to
$y$-paths.
Applying it to $P_{\to}(\mathbf{x}_{1:T})$ yields $P_{\to}(y_{1:T})$,
and applying it to $P_{\leftarrow}(\mathbf{x}_{T:1})$ yields $P_{\leftarrow}(y_{T:1})$.
The monotonicity of KL divergence under Markovian maps
(data-processing inequality) \cite{CoverThomas2006} therefore gives
\begin{equation}\label{eq:Sigma-full-geq-ME}
  \mathcal{S}_{\mathbf{x}}
   \geq 
  \mathcal{S}_y
   \geq  0.
\end{equation}
This is the coarse-graining inequality familiar from the
stochastic thermodynamics literature
\cite{RoldanParrondo2010,RoldanParrondo2012,GomezMarinParrondoVandenBroeck2008,KawaguchiNakayama2013,KahlenEhrich2018,CrooksStill2019}:
the entropy production of the Markovian process $\mathbf{x}_t$ is at
least as large as the irreversibility detectable from the observation
$y_t$ alone.

\subsection{Relation to the present framework}

We examine how the autoregressive framework of the main text relates to the Markovian embedding defined above.  
First, remember that in the general setting including Transformers,
$h_t = \Phi_t(y_1,\dots,y_t)$ does not factor through a two-argument
recursion, and thus even $(h_t, y_t)$ is not Markovian in general.

In the recursive case discussed in
Section~\ref{sec:recursive-special-case} with $h_t = \phi (h_{t-1}, y_t)$, the joint process $(h_t, y_t)$ is
Markovian. Therefore,  $\mathbf{x}_t \equiv (h_t, y_t)$ is a Markovian embedding of $y_t$.
(Note that  $h_t$ alone does not constitute
a Markovian embedding of $y_t$ in general.)

In this case, the forward path probability $P_{\to}(y_{1:T})$ defined in Section~\ref{sec:setup} coincides with the $y$-marginal of the joint forward process \eqref{eq:markov-embedding-def}.
Explicitly, we set
\begin{equation}\label{eq:ME-forward-kernel}
  p_t(\mathbf{x}_{t+1} \mid \mathbf{x}_t)
  = p_t(y_{t+1} \mid h_t)
    \delta \bigl(h_{t+1} - \phi (h_t,\, y_{t+1})\bigr);
\end{equation}
the recursive integration like 
\begin{equation}
\begin{split}
&\int dh_{t+1}  dh_t  p_t(y_{t+1} | h_t)  \delta(h_{t+1} - \phi (h_t, y_{t+1})) \\
&\quad \quad \quad \quad \quad \quad \times p_{t-1}(y_t | h_{t-1})  \delta(h_t - \phi (h_{t-1}, y_t)) \\
&= \int dh_t \, p_t(y_{t+1} | h_t)
  p_{t-1}(y_t | h_{t-1})  \delta(h_t - \phi (h_{t-1}, y_t)) \\
&= p_t(y_{t+1} | \phi (h_{t-1}, y_t))   p_{t-1}(y_t | h_{t-1})
 \end{split}
\end{equation}
confirms that it produces  $P_{\to}(y_{1:T})$ defined in Section~\ref{sec:setup}.
Moreover, the backward path probability $P_{\leftarrow}(y_{T:1})$ defined in \eqref{eq:backward-ME} above and that in Section~\ref{sec:backward} coincide,
if we choose the appropriate boundary term
\begin{equation}
\tilde p (\tilde{\mathbf{x}}_1) =\tilde p (\tilde h_1, \tilde y_1) \equiv \tilde{p}_0 (\tilde{y}_1 | \tilde{h}_0) \delta ( \tilde{h}_1 - \phi (\tilde{h}_0, \tilde{y}_1) ),
\end{equation}
where $\tilde h_0 = h_0$ is a fixed constant as in our general framework.
Indeed, by letting $\tilde y_s = y_{T-s+1}$, a similar recursive integration gives
\begin{equation}
\begin{split}
&\int d\tilde{h}_{s+1}  d\tilde{h}_s  p_{T-s}(y_{T-s} | \tilde{h}_s)  \delta(\tilde{h}_{s+1} -  \phi (\tilde{h}_s, y_{T-s})) \\
&\quad \quad \quad \times  p_{T-s+1}(y_{T-s+1} | \tilde{h}_{s-1})  \delta(\tilde{h}_s -\phi(\tilde{h}_{s-1}, y_{T-s+1})) \\
&= \int d\tilde{h}_s  p_{T-s}(y_{T-s} | \tilde{h}_s)   p_{T-s+1}(y_{T-s+1} | \tilde{h}_{s-1})  \\
&\quad \quad \quad \quad \quad \quad \quad \times \delta(\tilde{h}_s -  \phi(\tilde{h}_{s-1}, y_{T-s+1})) \\
&= p_{T-s}(y_{T-s} | \phi (\tilde{h}_{s-1}, y_{T-s+1})) p_{T-s+1}(y_{T-s+1} | \tilde{h}_{s-1}).
  \end{split}
\end{equation}
Therefore,  $\mathcal{S}_y$ defined above coincides with our general definition of the entropy production \eqref{eq:Sigma-def}.

On the other hand, as noted in Section~\ref{sec:backward_ep},  applying the deterministic maps $\phi$ backwards to the time-reversed sequence $y_{T:1} $ does not reproduce the reversed sequence of $h_{1:T}$ in general; $\tilde h_{s} = h_{T-s+1}$ does not necessarily hold.
Explicitly, if we substitute $\tilde{\mathbf{x}}_{1:T} = \mathbf{x}_{T:1}$ into the backward path probability \eqref{eq:backward-ME}, each transition kernel becomes 
\begin{equation}\label{eq:ME-backward-kernel}
  p_t(\mathbf{x}_{t} \mid \mathbf{x}_{t+1})
  = p_t(y_{t} \mid h_{t+1})\,
    \delta \bigl(h_{t} - \phi (h_{t+1},\, y_{t})\bigr).
\end{equation}
In general,  the arguments of the delta functions in \eqref{eq:ME-forward-kernel} and \eqref{eq:ME-backward-kernel} cannot be zero at the same time.
As a consequence, $\tilde{\mathbf{x}}_{1:T} = \mathbf{x}_{T:1}$ is not realized in general,  implying that the deterministic part of the dynamics may be completely irreversible. Therefore, $\mathcal{S}_{\mathbf{x}}$ defined in \eqref{eq:Sigma-x-def} may diverge and would not be informative.
This consideration provides another basis for our approach to the non-Markovian entropy production, which is defined solely by the path probabilities of $y_{1:T}$ without embedding into another Markovian dynamics.

\subsection{True environmental state as a possible Markovian embedding}

Behind the observations $y_t$ there may exist a true environmental
state $x_t$ whose dynamics generates $y_t$.
If the joint process $(x_t, y_t)$ satisfies the factorization
\eqref{eq:markov-embedding-def} with
$\mathbf{x}_t = x_t$, then $x_t$ constitutes a Markovian embedding
of $y_t$.
The Kalman filter example
(Section~\ref{sec:linear-gaussian-example}) is a concrete instance:
$x_t$ evolves as $x_{t+1} = A x_t + w_t$ and produces observations
$y_t = C x_t + v_t$, which is exactly of the form
\eqref{eq:markov-embedding-def}.

In the present general framework, however, a true environmental state $x_t$
is not explicitly assumed.
The latent state $h_t = \Phi_t(y_1,\dots,y_t)$ is a deterministic
function of the observations and carries no independent stochastic
degrees of freedom; it is not an environmental state but a
computational latent state.

%======================================================================
\section{Details of the GPT-2 experiment}
\label{sec:suppl}
%======================================================================

We describe the details of the GPT-2 experiment
in Section~\ref{sec:gpt2-demo} and show some supplemental results.
Our implementation uses the Hugging Face
\texttt{transformers} library~\cite{Wolf2020},
which provides pre-trained weights and a tokenizer
for GPT-2.

The Hugging Face implementation may differ
from the original OpenAI release~\cite{Radford2019}
in default generation parameters and tokenizer behavior;
the details relevant to our experiment are described
in the subsections below.
Note that the original paper~\cite{Radford2019} reported 117M parameters for GPT-2;
OpenAI later noted that this count was erroneous~\cite{gpt2repo}, and
\texttt{model.parameters()} in Hugging Face yields approximately 124M, as in our metadata.

All GPT-2 sampling and likelihood-evaluation runs were performed on NVIDIA T4 or V100 GPUs. Exact token-by-token reproducibility of sampled trajectories is not guaranteed across different hardware/software environments, even with a fixed random seed.

\subsection{Details of sampling from GPT-2}
\label{app:gpt2-boundary}

For the sampling experiment from GPT-2 itself (Section~\ref{subsec:gpt2-generated}), the path probability in
Eq.~\eqref{eq:P-homogeneous} contains the boundary term
$p(y_1 \mid h_0)$.
This term must be specified separately, because the tokenizer
used in our implementation does not prepend an initial
beginning-of-sequence (BOS) token automatically.
In the Hugging Face GPT-2 tokenizer, the EOS token and the BOS token 
are both assigned to the same token \texttt{<|endoftext|>},
and the tokenizer backend adds the BOS token only when explicitly requested~\cite{gpt2tokenizercode,tokenizersbackendcode}.

Accordingly, our implementation explicitly prepends
\texttt{<|endoftext|>} to the tokenized sequence before each
forward pass: the model receives
$[\texttt{<|endoftext|>}, y_1, \dots, y_T]$.
Therefore, 
\begin{equation}
P(y_{1:T}) = \prod_{t=0}^{T-1} p(y_{t+1} \mid \texttt{<|endoftext|>}, y_1, \dots, y_t);
\end{equation}
the map $\Phi(y_1, \dots, y_t)$
 corresponds to the latent state after processing
$[\texttt{<|endoftext|>}, y_1, \dots, y_t]$.
Note that GPT-2 uses absolute position embeddings.
For $t=0$, we adopt the initial condition
\begin{equation}
  p(y_1 \mid h_0)
  \equiv
  p \left(y_1 \mid \texttt{<|endoftext|>}\right).
\end{equation}
Since \texttt{<|endoftext|>} is always fixed, the above construction exactly fits our general theoretical framework.

Hugging Face has a default function \texttt{model.generate()} for generating tokens.
In our experiment, to make the construction more explicit, we replace
\texttt{model.generate()} with an explicit autoregressive loop
that uses the KV-cache of GPT-2 directly.
At each step~$t$, the logits are computed from the cached
latent state, the next token~$y_{t+1}$ is drawn from
$p(\cdot \mid \texttt{<|endoftext|>}, y_1, \dots, y_t)$,
and the log-likelihood
$\ln p(y_{t+1} \mid \texttt{<|endoftext|>}, y_1, \dots, y_t)$
is recorded from the same logits.
Therefore, in our implementation, the forward log-likelihood is taken from
exactly the same logits within each run.

Meanwhile, Hugging Face's \texttt{model.generate()} terminates decoding upon emitting the EOS token
\texttt{<|endoftext|>}.
On the other hand, our above implementation  always executes exactly $T$~steps of token generation, where
the EOS token may appear within the generated sequence
but does not trigger termination.

For the reversed sequence, we use
$[\texttt{<|endoftext|>}, y_T, y_{T-1}, \dots, y_{1}]$ to calculate the backward log-likelihood,
so that the boundary term is treated
identically in both directions.
That is, the same fixed initial context is used for $y_{1:T}$ and for
$y_{T:1}$, and the assumption $\tilde{h}_0 = h_0$ is satisfied.
The reverse log-likelihood
$\ln P(y_{T}, y_{T-1}, \dots, y_{1})$ is obtained by
feeding the reversed sequence
$[\texttt{<|endoftext|>}, y_T, y_{T-1}, \dots, y_1]$
to the model by using the same protocol as the forward process with the KV-cache.

We do not need any stronger statement
about the unpublished details of the original training-data
pipeline.
For the numerical experiment, the only required point is that
the rule above gives a concrete and reproducible definition of
the boundary term for the publicly released model.

Note that, in the fixed-text experiment of Section~\ref{subsec:gpt2-fixed} where no sampling is performed, the log-likelihood of each sequence is evaluated by a single full-sequence forward pass of GPT-2 (one pass per sequence), rather than by the KV-cache incremental path used in Section~\ref{subsec:gpt2-generated}.

\subsection{Convergence of entropy production and fluctuation theorem}
\label{app:gpt2-ft}

We next show supplemental numerical results for the Monte Carlo sampling from GPT-2.

\begin{figure*}[htbp]
  \centering
  \includegraphics[width=0.8\linewidth]{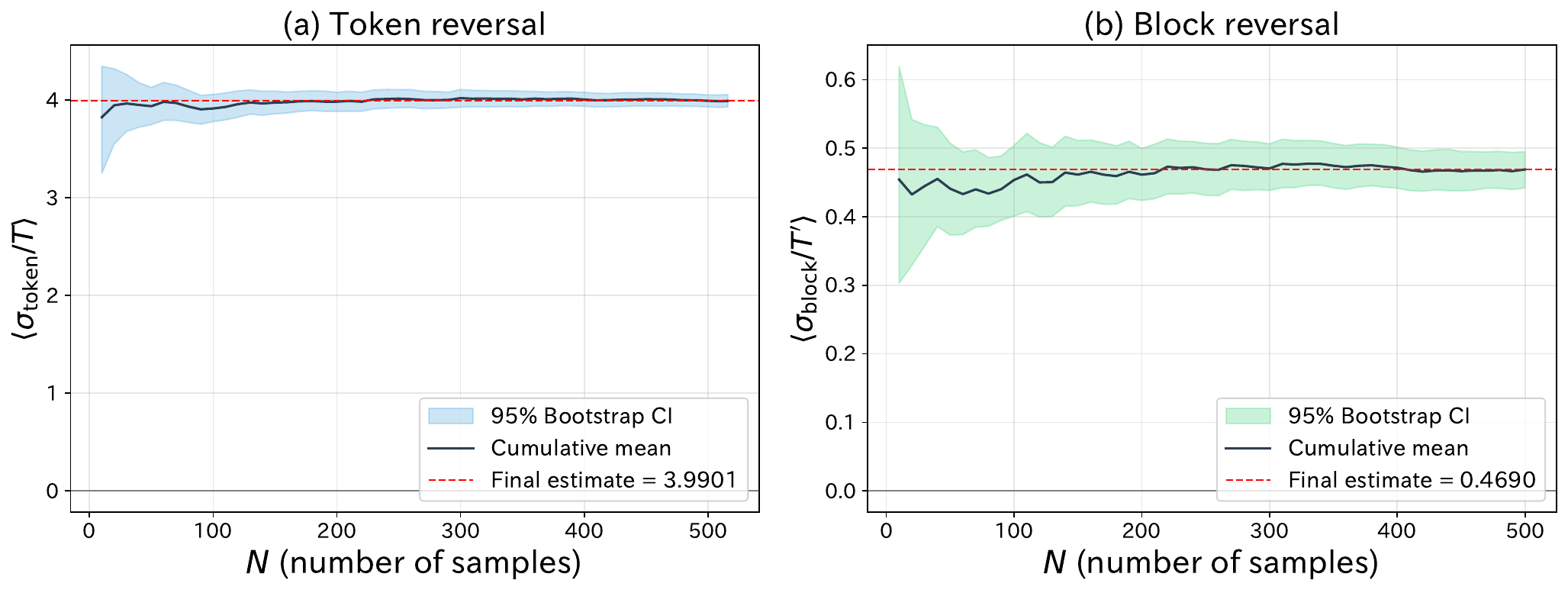}
  \caption{Convergence of the Monte Carlo estimates
  of the per-token entropy production for $T=120$ in GPT-2,
  as a function of the number of samples~$N$.
  The temperature parameter is $\tau = 1$.
    (a)~Token-level reversal
  $\sigma_{\mathrm{token}}/T$;
  (b)~block-level reversal
  $\sigma_{\mathrm{block}}/T'$.
  Solid lines show the cumulative sample mean;
  shaded regions are $95\%$ bootstrap percentile
  confidence intervals ($B = 2000$).
  Dashed red lines indicate the final estimates at $N = 516$ or $500$ (see the caption of Figure~\ref{fig:gpt2-histogram}).}
  \label{fig:gpt2-convergence}
\end{figure*}

To examine how the Monte Carlo estimates stabilize
as the sample size grows,
Figure~\ref{fig:gpt2-convergence} plots the cumulative
sample mean of  $\sigma_{\mathrm{token}}/T$ and
$\sigma_{\mathrm{block}}/T'$ for $T=120$, as a function of~$N$.
The shaded bands indicate $95\%$ confidence intervals
obtained by the bootstrap percentile method
($B = 2000$ resamples at each~$N$),
which does not assume normality of the sampling distribution.
Both estimators converge to positive values,
consistent with the positivity of the entropy production.

\begin{figure*}[htbp]
  \centering
  \includegraphics[width=0.8\linewidth]{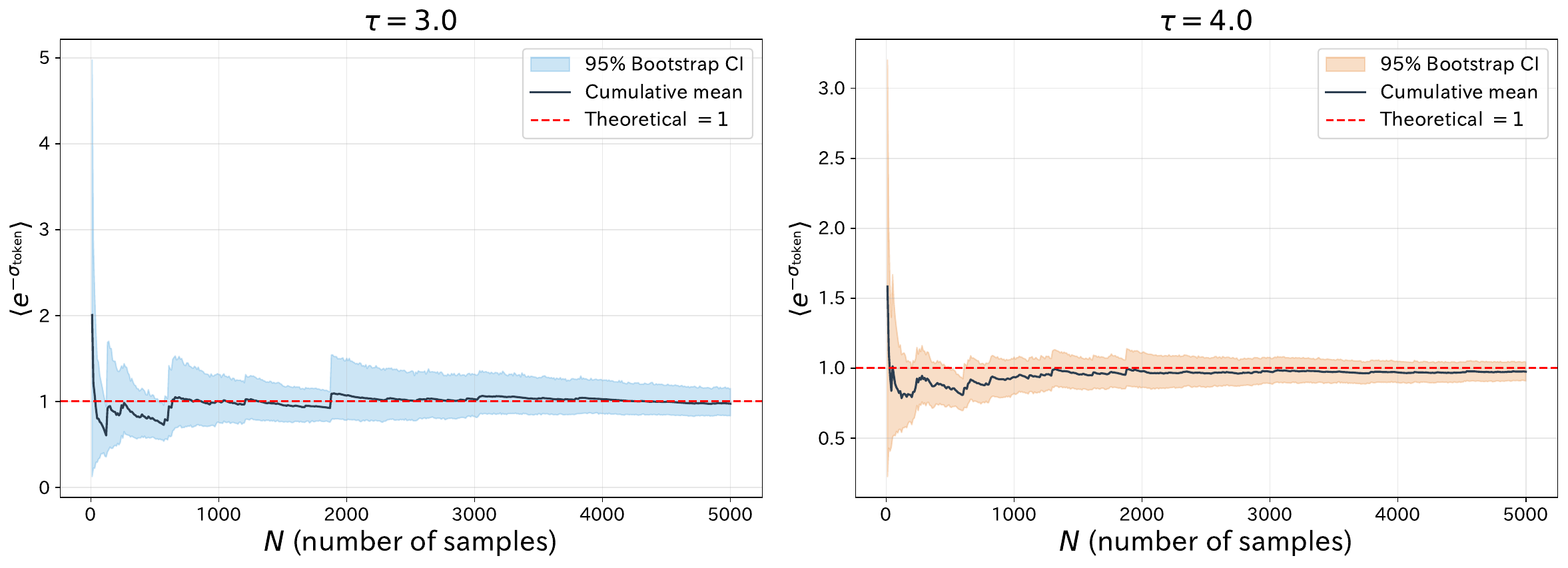}
  \caption{Convergence of the Monte Carlo mean of
  $e^{-\sigma_{\mathrm{token}}}$
  as a function of the number of samples~$N$,
  for sequences of $T = 5$ tokens sampled from GPT-2.
  Temperatures are $\tau = 3$ and $\tau = 4$.
  Solid lines show the cumulative sample mean;
  shaded regions are $95\%$ bootstrap percentile
  confidence intervals ($B = 2000$).
  The dashed red line indicates the theoretical
  prediction
  $\mathbb{E}_P [ e^{-\sigma_{\mathrm{token}}} ] = 1$, namely the integral fluctuation theorem.
  In both panels the cumulative mean trends toward
  the theoretical value $1$, as~$N$ grows.}
  \label{fig:gpt2-ft-convergence}
\end{figure*}

We also perform a numerical verification of the fluctuation theorem
$\mathbb{E}_P [ e^{-\sigma_{\mathrm{token}}} ] = 1$,
where $\sigma_{\mathrm{token}}$ is the stochastic entropy production with token-level reversal without truncation, and the samples are drawn from the  forward distribution of GPT-2. 
Because $\sigma_{\mathrm{token}}$ grows with $T$,
choosing a large sequence length makes
$e^{-\sigma_{\mathrm{token}}}$ extremely small for typical $\sigma_{\mathrm{token}} > 0$,
resulting in intractably slow convergence
of the exponential average.
We therefore set $T = 5$ in this experiment.
Similarly, since the convergence is poor at $\tau = 1$,
we instead adopt $\tau = 3$ and $\tau = 4$,
for which the cumulative sample mean roughly
converges towards the theoretical value of~$1$
as the number of samples~$N$ increases.
Figure~\ref{fig:gpt2-ft-convergence} shows this
convergence behavior for $N = 5000$ samples
at each temperature.
While the fluctuation theorem directly follows from the definition of the entropy production, our numerical result serves as a consistency check for our sampling method.

\subsection{Dependence on the block scale}
\label{app:block-scale-dependence}

We further examine how the entropy production depends on the scale of the
block reversal, using the same $N=500$ truncated sequences
$y_{1:T'}$ as in Fig.~\ref{fig:gpt2-histogram}.  This analysis reuses the saved
trajectories; no new text is sampled.  
For each original or reordered sequence, the log-likelihood is re-evaluated with GPT-2.

\begin{figure}[t]
  \centering
  \includegraphics[width=1.0\linewidth]{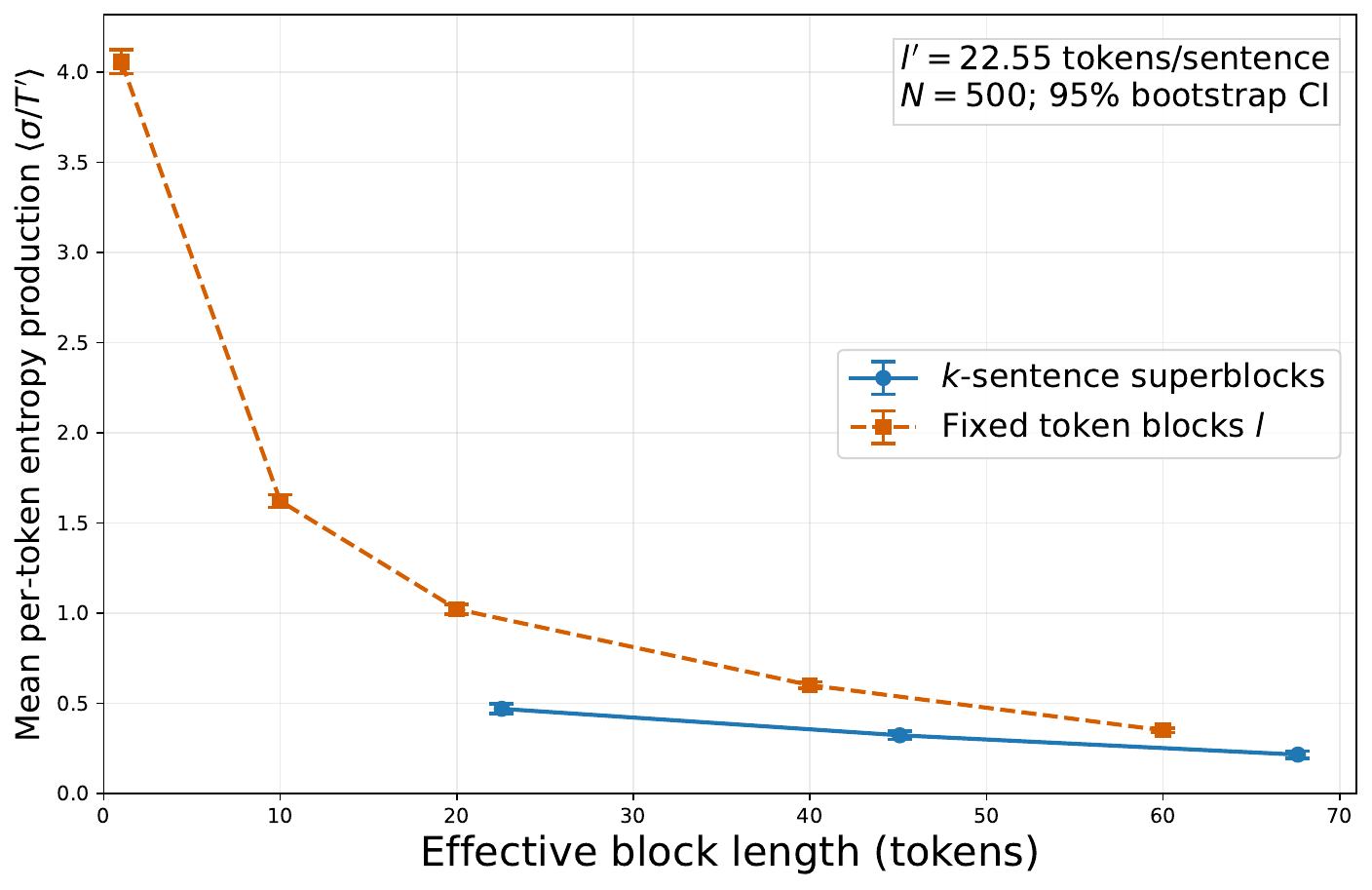}
  \caption{Dependence of the mean per-token entropy production on
  the block scale for the $N=500$ GPT-2 samples used in Fig.~\ref{fig:gpt2-histogram}.  The blue circles show reversals of superblocks formed from $k$ consecutive sentences and are plotted at $kl'$, with
  $l'=22.55$ tokens per sentence; the plotted values are $k=1,2,3$.
  The orange squares show reversals of fixed-length blocks with block length $l=1,10,20,40,60$ and are plotted at $l$.  Each point is the arithmetic   mean of the per-sample values divided by $T'$, and error bars show 95\%
  percentile bootstrap confidence intervals ($B=2000$).  Lines are guides to   the eye.}
  \label{fig:gpt2-block-scale}
\end{figure}

We consider two block constructions.  First, for the sentence-based
construction, sentence boundaries are reconstructed using the same
sentence-final punctuation-token rule as in Fig.~\ref{fig:gpt2-histogram} (b).  Consecutive groups of $k$ sentences are
treated as superblocks and the superblock order is reversed, while the order
of sentences and tokens within each superblock is retained.  The final
superblock is allowed to contain fewer than $k$ sentences.  
The blue circles in Fig.~\ref{fig:gpt2-block-scale} show the mean per-token entropy production for this sentence-based construction.
The horizontal
axis is $kl'$, where $l' = 22.55$ is the mean sentence length.
To focus on block scales up to approximately 60--70 tokens, we used $k=1,2,3$.  
The case $k=1$ corresponds to the sentence-block
reversal used in  Fig.~\ref{fig:gpt2-histogram} (b).

Second, we consider the fixed-length construction discussed in Section~\ref{sec:implications-cg}, where each sequence was split from its
beginning into consecutive blocks of $l$ tokens with $l$ being fixed.
Note that the final block was allowed to be shorter than $l$.  
The orange squares of  Fig.~\ref{fig:gpt2-block-scale} show the mean per-token entropy production
for this fixed-length construction.
We use $l=1,10,20,40,60$, and plot each result at $l$ on the horizontal axis.  In particular, $l=1$ corresponds to the token-level reversal of the same truncated sequences $y_{1:T'}$.

As seen from Fig.~\ref{fig:gpt2-block-scale}, the mean entropy production
decreases over the plotted block-scale range for both constructions.  
The entropy production for the fixed-length construction is always greater than that for the sentence-based construction. This is consistent with the observation that the former contains syntactic artifacts, because block boundaries do not coincide with sentence boundaries.

\subsection{Input text sets and statistical analysis}
\label{app:fixed-texts}

For the fixed-text experiment in Section~\ref{subsec:gpt2-fixed}, we used five language
models as input-text generators: Claude Opus 4.6, Claude Fable 5, Gemini 3.1 Pro, GPT-5.4 Pro, and GPT-5.6 Pro.  The same fixed prompt
was submitted to each generator.  For every generator, the resulting
corpus contains 100 causal and 100 non-causal English texts, with four
sentences in each text.  The causal texts are short narratives in
which the sentences describe a temporally ordered sequence of events,
whereas the non-causal texts consist of independent statements whose
ordering has no intended temporal or causal implication.  These
categories are defined operationally by the prompt and are not based
on a formal criterion for causal dependence.  The complete prompts,
text sets, analysis code, and raw numerical results are provided in
the GitHub repository (see the Data Availability statement).

The prompt contains one causal example and one non-causal example,
both of which are required to appear as fixed entries in the generated
lists.  Consequently, these two texts are shared across all five
corpora, while the remaining 99 causal and 99 non-causal texts are
generator-specific.  The shared examples are the two texts displayed
in Section~\ref{subsec:gpt2-fixed}.
When a generator's output violated the required format (e.g., invalid JSON or an incorrect number of texts or sentences), the generation was repeated with the same fixed prompt until the format constraints were satisfied; the resulting texts were used verbatim, with no editing and no selection based on their content.
Note that three of the 100 causal texts generated by Gemini 3.1 Pro contain ``then'' or ``later'' despite the prompt's prohibition of such words.
We note that each corpus was obtained from one model response, and therefore, strictly speaking, the texts within a corpus are not guaranteed to be statistically independent.

\begin{table*}[t]
  \caption{Statistical comparison of the GPT-2 experiment with causal and non-causal fixed-text
  samples discussed in Section~\ref{subsec:gpt2-fixed}.
  For every row, $n_{\mathrm{C}}=100$ and $n_{\mathrm{NC}}=100$, and all
  observations are included without outlier removal.
  $U_{\mathrm{C}}$ is the Mann--Whitney statistic for the causal
  sample, and
  $r=2U_{\mathrm{C}}/(n_{\mathrm{C}}n_{\mathrm{NC}})-1$ is the
  rank-biserial correlation; $r>0$ indicates that the causal values
  tend to be larger than the non-causal values.
  The reported $p$ values are from the individual two-sided
  asymptotic Mann--Whitney test.}
  \label{tab:fixed-text-statistics}
  \begin{ruledtabular}
    \begin{tabular}{llrcc}
      Input-text generator & Reversal & $U_{\mathrm{C}}$ & $r$ & $p$ \\
      \hline
      Claude Opus 4.6 & Token & 3282 & $-0.3436$ & $2.710\times10^{-5}$ \\
      Claude Opus 4.6 & Block & 7286 & $0.4572$ & $2.346\times10^{-8}$ \\
      Claude Fable 5 & Token & 3984 & $-0.2032$ & $0.01309$ \\
      Claude Fable 5 & Block & 6356 & $0.2712$ & $9.263\times10^{-4}$ \\
      Gemini 3.1 Pro & Token & 5507 & $0.1014$ & $0.2159$ \\
      Gemini 3.1 Pro & Block & 6312 & $0.2624$ & $0.001353$ \\
      GPT-5.4 Pro & Token & 5389 & $0.0778$ & $0.3425$ \\
      GPT-5.4 Pro & Block & 5467 & $0.0934$ & $0.2544$ \\
      GPT-5.6 Pro & Token & 2687 & $-0.4626$ & $1.601\times10^{-8}$ \\
      GPT-5.6 Pro & Block & 5468 & $0.0936$ & $0.2533$ \\
    \end{tabular}
  \end{ruledtabular}
\end{table*}

All likelihoods are evaluated using GPT-2 at temperature $\tau = 1$.
To preserve inter-block whitespace when sentence blocks are permuted at the token-ID level, each fixed text is tokenized with the GPT-2 tokenizer using \texttt{add\_prefix\_space=True}; this encodes the first word in the same space-prefixed form as words preceded by whitespace elsewhere in the text.
As in Appendix B 1, every original or reversed sequence is evaluated with the fixed \texttt{<|endoftext|>} token as its initial context.

For each input-text generator and reversal scheme, we compare the
causal and non-causal samples using a two-sided asymptotic
Mann--Whitney $U$ test with the standard tie correction to the
variance and with continuity correction.  
The direction and magnitude of each difference are
quantified by the rank-biserial correlation $r$ defined in
Section~\ref{subsec:gpt2-fixed}.  Pointwise 95\% confidence intervals for $r$ are obtained
using the percentile bootstrap with independent resampling within the
causal and non-causal categories.
The summary of these test statistics is shown in Table~\ref{tab:fixed-text-statistics}.
All reported $p$ values are individual and unadjusted for multiplicity.

\section{Fixed-text analysis with Qwen3-4B-Base}
\label{app:qwen3-fixed-text}

In this appendix, we examine whether the fixed-text results of  Section~\ref{subsec:gpt2-fixed} (Appendix~\ref{app:fixed-texts}) depend on the relatively small GPT-2 evaluator, by repeating the entire analysis with a substantially larger evaluator, Qwen3-4B-Base~\cite{qwen3technicalreport,qwen3base-modelcard}.
Because the input corpora are unchanged, this analysis tests robustness across evaluator models and tokenizers.

Qwen3-4B-Base is an open-weight, pre-trained language model with
4.0 billion parameters.  We used the base model rather than a post-trained instruction or
chat model.  Accordingly, no chat template, system prompt, or generation
prompt was applied.
All likelihood-evaluation runs were performed on NVIDIA V100 GPUs. 

The evaluator was applied separately to the same five fixed-text corpora
used in Section~\ref{subsec:gpt2-fixed} (Appendix~\ref{app:fixed-texts}). 
For likelihood evaluation, one ASCII 
space was prepended when a string did not already begin with a space, and
the resulting string was tokenized by the Qwen3 tokenizer with
\texttt{add\_special\_tokens=False}.  This explicit leading-space rule
plays the role of \texttt{add\_prefix\_space=True} in the GPT-2
implementation and preserves the whitespace attached to sentence blocks
when their token-ID sequences are permuted.  
No EOS target was appended; $T$ counts only the tokens obtained from the fixed input text.
Reversal was performed
only after tokenization; reversed strings were not re-tokenized.

As in the GPT-2 analysis, we used an explicit fixed initial-context token, although the relevant tokenizer metadata differs.
 The Qwen3 tokenizer does not automatically prepend a BOS token; we explicitly chose \texttt{model.config.bos\_token\_id}, namely token ID 151643 (\texttt{<|endoftext|>}), as the common fixed initial-context token for the original and reversed paths. The official model configuration designates this token as both BOS and EOS, and it coincides with the tokenizer's EOS token. Thus, the resulting boundary term parallels the \texttt{<|endoftext|>} convention of the GPT-2 implementation (Appendix~\ref{app:gpt2-boundary}). 

For each text, the forward log-likelihood was evaluated once and reused
in the two reversal analyses.  
  Likelihood evaluation used \texttt{torch.inference\_mode()} and
\texttt{use\_cache=False}.  
We report $\sigma_{\mathrm{token}}/T$ and
$\sigma_{\mathrm{block}}/T$, where $T$ is the number of tokens assigned by
the Qwen3 tokenizer.  The GPT-2 and Qwen3 tokenizers generally
produce different token sequences and different values of $T$.

\begin{figure*}[t]
  \centering
  \includegraphics[width=\textwidth]{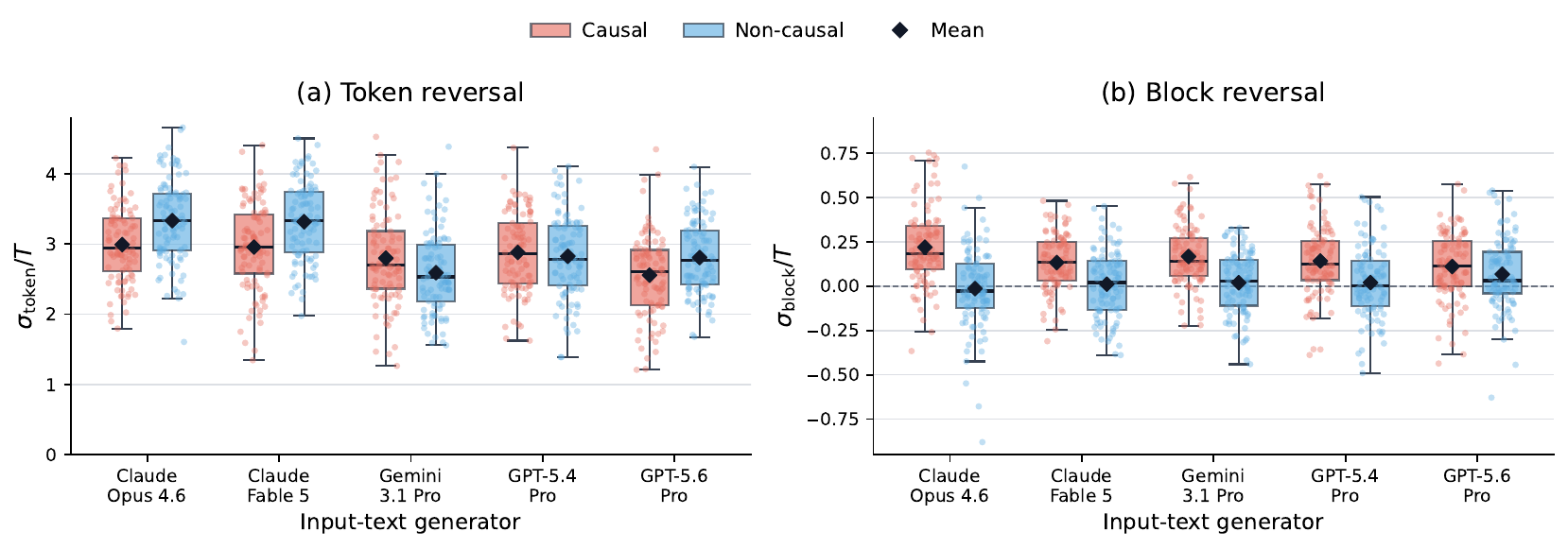}
  \caption{Distributions of the stochastic entropy production evaluated
  using Qwen3-4B-Base for the fixed texts generated by the five input-text
  models.  (a) Token-level reversal $\sigma_{\mathrm{token}}/T$.
  (b) Sentence-block reversal $\sigma_{\mathrm{block}}/T$. The plotting conventions are the same as those in Fig.~\ref{fig:fixed-text-distributions}.}
  \label{fig:qwen3-fixed-text-distributions}
\end{figure*}

\begin{figure*}[t]
  \centering
  \includegraphics[width=\textwidth]{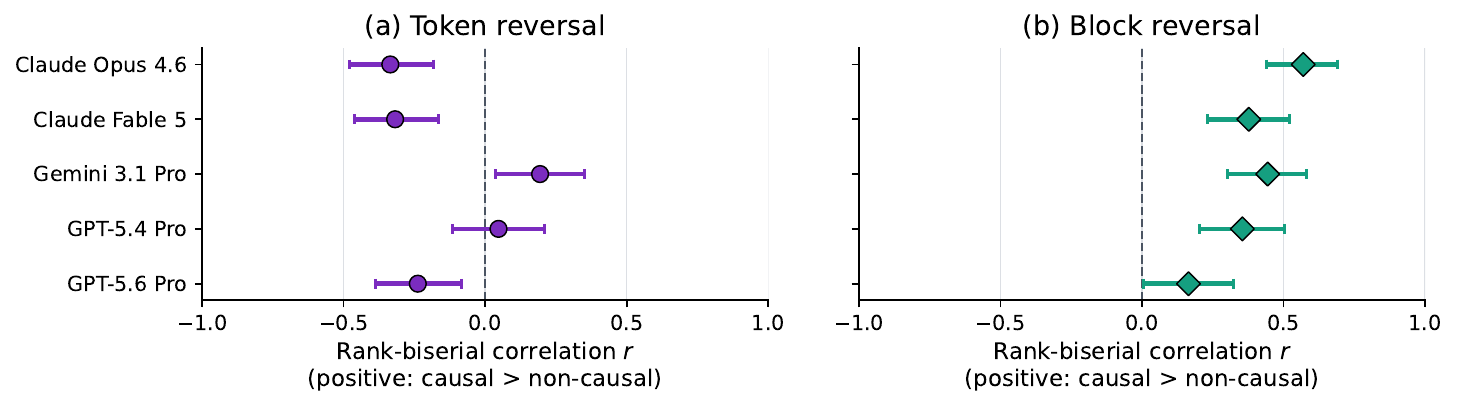}
  \caption{Cross-generator comparison of the causal/non-causal difference
  in the per-token stochastic entropy production evaluated by
  Qwen3-4B-Base.  The labels on the vertical axis indicate the language
  model used to generate each input-text corpus.  (a) Token-level reversal
  $\sigma_{\mathrm{token}}/T$.  (b) Sentence-block reversal
  $\sigma_{\mathrm{block}}/T$.  The plotting conventions and statistical annotations are the same as those in Fig.~\ref{fig:fixed-text-cross-model}.}
  \label{fig:qwen3-fixed-text-effects}
\end{figure*}

The resulting distributions are shown in
Fig.~\ref{fig:qwen3-fixed-text-distributions}, and the corresponding
rank-biserial correlations are shown in
Fig.~\ref{fig:qwen3-fixed-text-effects}.  
The summary of the test statistics is reported in Table~\ref{tab:fixed-text-statistics2}.

The qualitative pattern closely reproduces the GPT-2 result of
Section~\ref{subsec:gpt2-fixed}, and the statistical resolution
improves. For token reversal, the causal/non-causal difference again
depends strongly on the input-text generator, with significant
differences now occurring in both directions: significantly smaller
$\sigma_{\mathrm{token}}/T$ for the causal texts of Claude Opus 4.6,
Claude Fable 5, and GPT-5.6 Pro, significantly
larger for those of Gemini 3.1 Pro, and no
significant difference for GPT-5.4 Pro. Thus, even with the
substantially larger Qwen3 evaluator, the token-level direction remains
a generator-specific property of the text corpora.

For block reversal, by contrast, the causal texts show significantly
larger $\sigma_{\mathrm{block}}/T$ for all five input-text generators, including GPT-5.4 Pro and GPT-5.6 Pro, for which no
significant difference was resolved with GPT-2. The same direction is
therefore recovered across the five input-text generators and the two
evaluators, despite the large difference in model size and the use of
different tokenizers. This provides additional support for the
conclusion of Section~\ref{subsec:gpt2-fixed} that block reversal is
the more suitable probe of inter-sentence ordering.
As before, however, it does
not by itself establish $\sigma_{\mathrm{block}}/T$ as a general
quantitative measure of causal structure.

\begin{table*}[t]
  \caption{Statistical comparison of the Qwen3-4B-Base experiment with causal and non-causal fixed-text   samples.  The statistical annotations are the same as those in Table~\ref{tab:fixed-text-statistics}.}
  \label{tab:fixed-text-statistics2}
  \begin{ruledtabular}
    \begin{tabular}{llrcc}
      Input-text generator & Reversal & $U_{\mathrm{C}}$ & $r$ & $p$ \\
      \hline
      Claude Opus 4.6 & Token & 3324 & $-0.3352$ & $4.242\times10^{-5}$ \\
      Claude Opus 4.6 & Block & 7846 & $0.5692$ & $3.585\times10^{-12}$ \\
      Claude Fable 5 & Token & 3409 & $-0.3182$ & $1.018\times10^{-4}$ \\
      Claude Fable 5 & Block & 6886 & $0.3772$ & $4.085\times10^{-6}$ \\
      Gemini 3.1 Pro & Token & 5969 & $0.1938$ & $0.01796$ \\
      Gemini 3.1 Pro & Block & 7219 & $0.4438$ & $5.938\times10^{-8}$ \\
      GPT-5.4 Pro & Token & 5235 & $0.0470$ & $0.5667$ \\
      GPT-5.4 Pro & Block & 6771 & $0.3542$ & $1.518\times10^{-5}$ \\
      GPT-5.6 Pro & Token & 3810 & $-0.2380$ & $0.003656$ \\
      GPT-5.6 Pro & Block & 5820 & $0.1640$ & $0.04525$ \\
    \end{tabular}
  \end{ruledtabular}
\end{table*}

 \FloatBarrier

%======================================================================
%======================================================================

\end{document}